\documentclass[10pt,american,english,authoryear,3p]{elsarticle}
\usepackage[T1]{fontenc}
\usepackage[latin9]{inputenc}
\usepackage{textcomp}
\usepackage{algorithm2e}
\usepackage{amsmath}
\usepackage{amsthm}
\usepackage{amssymb}
\usepackage{graphicx}

\makeatletter

\providecommand{\tabularnewline}{\\}

\makeatother

\usepackage{babel}
\begin{document}

\begin{frontmatter}{}

\title{CellEVAC: An adaptive guidance system for crowd evacuation through
behavioral optimization}

\author[1]{Miguel A. Lopez-Carmona\corref{cor1}} \ead{miguelangel.lopez@uah.es}
\author[1]{Alvaro Paricio Garcia} \ead{alvaro.paricio@uah.es}
\cortext[cor1]{Corresponding author} \address[1]{Universidad de Alcala, Escuela Politecnica Superior, Departamento de Automatica, Campus Externo de la Universidad de Alcala, Alcala de Henares, Madrid, Spain}
\begin{abstract}
A critical aspect of crowds' evacuation processes is the dynamism
of individual decision making. Identifying optimal strategies at an
individual level may improve both evacuation time and safety, which
is essential for developing efficient evacuation systems. Here, we
investigate how to favor a coordinated group dynamic through optimal
exit-choice instructions using behavioral strategy optimization. We
propose and evaluate an adaptive guidance system (Cell-based Crowd
Evacuation, CellEVAC) that dynamically allocates colors to cells in
a cell-based pedestrian positioning infrastructure, to provide efficient
exit-choice indications. The operational module of CellEVAC implements
an optimized discrete-choice model that integrates the influential
factors that would make evacuees adapt their exit choice. To optimize
the model, we used a simulation-optimization modeling framework that
integrates microscopic pedestrian simulation based on the classical
Social Force Model. In the majority of studies, the objective has
been to optimize evacuation time. In contrast, we paid particular
attention to safety by using Pedestrian Fundamental Diagrams that
model the dynamics of the exit gates. CellEVAC has been tested in
a simulated real scenario (Madrid Arena) under different external
pedestrian flow patterns that simulate complex pedestrian interactions.
Results showed that CellEVAC outperforms evacuation processes in which
the system is not used, with an exponential improvement as interactions
become complex. We compared our system with an existing approach based
on Cartesian Genetic Programming. Our system exhibited a better overall
performance in terms of safety, evacuation time, and the number of
revisions of exit-choice decisions. Further analyses also revealed
that Cartesian Genetic Programming generates less natural pedestrian
reactions and movements than CellEVAC. The fact that the decision
logic module is built upon a behavioral model seems to favor a more
natural and effective response. We also found that our proposal has
a positive influence on evacuations even for a low compliance rate
(40\%).
\end{abstract}

\end{frontmatter}{}

\section{Introduction}

Destructive and uncoordinated crowd behaviors such as herding or stampede
are recognized as being responsible for pedestrians' death and injury
in large-scale crowd evacuations during emergencies. Evacuees tend
to seek their safety and exhibit selfish attitudes that may go against
the collective benefit. An efficient evacuation plan is of paramount
importance to coordinate and direct evacuees out of dangerous areas
in a safe and timely manner. This coordination can be achieved by
deploying guidance systems capable of providing information for each
user on the exit gate, the path to follow, and possibly the time when
evacuation should start \citep{Abdelghany20141105}. These systems
may embed real-time routing algorithms that provide adaptive plans
or use pre-deployed static plans based on prediction and analysis
\citep{biSurveyAlgorithmsSystems2019}.

It is well known that the performance of evacuation processes can
be strongly affected by exit-choice decision making at the individual
level \citep{haghaniSimulatingDynamicsAdaptive2019}. Research on
human responses to multiple sources of directional information has
shown that directional information has a significant effect on human
exit route choice \citep{bodeHumanResponsesMultiple2014}. For instance,
\citep{felicianiEfficientlyInformingCrowds2020} studied how egress
time can be reduced if wheelchair users are informed on exit type
and location. Thus, there are research efforts in the area of real-time
routing for crowd evacuations that have focused on studying mechanisms
for providing optimal exit-choice information. These mechanisms have
been mainly implemented using optimal static plans obtained through
simulation-optimization methods \citep{Abdelghany20141105,guoPotentialbasedDynamicPedestrian2018}.
However, since the dynamics of the environment change over time in
unpredictable ways during emergency evacuations, adaptive strategies
appear more appealing \citep{Zhong2016127}.

Developing evacuation systems based on adaptive exit-choice information
is challenging. For example, human response to information given by
the system during emergencies should be easy to understand and follow,
but most existing research addresses the algorithmic perspective and
overlook the systemic view and usability. We consider this aspect
as essential if we want to take these systems to real environments.
When optimizing evacuation plans in general, the overall objective
has been minimizing the total evacuation time, average evacuation
time, or maximizing the cumulative exit throughput within a given
period. Surprisingly, the safety dimension has not been explicitly
considered, being considered a consequence of applying different optimization
objectives on the facility configuration and demand profiles. We believe
that safety should be an explicit objective when optimizing evacuation
plans because evacuation time and safety dimensions are closely related
to each other in emergency evacuations and under extreme conditions.
Finally, though it is widely accepted that the modeling of evacuation
behavior is essential for developing efficient evacuation systems,
the primary attention has been being paid to investigate the problem
from architectural or path-planning optimization perspectives \citep{duivesStateoftheartCrowdMotion2013,vermuytenReviewOptimisationModels2016,zhaoOptimalLayoutDesign2017}.
The potential of behavioral strategy optimization for improving both
evacuation time and safety has been overlooked in the design of adaptive
evacuation systems.

We are interested in quantifying the benefits of using evacuation
behavior models to implement adaptive evacuation systems' decision
logic. Another central question of this study concerns the topic of
safety. We would like to quantify how important it is to explicitly
model safety, and which could be a reasonable safety metric.

With the purposes mentioned above, this paper proposes an adaptive
guidance system (Cell-based Crowd Evacuation, CellEVAC) that dynamically
maps exit-choice indications in a cell-based positioning infrastructure.
During the evacuation process, colors (indications, instructions)
corresponding to an exit gate are dynamically displayed in pedestrians'
wearable or personal handheld devices depending on the current pedestrian
cell-position. The kernel of the system is a Multinomial Logit Model
(MLM) taken from discrete choice theory, which has been widely used
to model human behavior in evacuations and many other areas such as
economics or transportation \citep{pressDiscreteChoiceAnalysis1985,duivesExitChoiceDecisions2012}.
In our study, we use it to implement the decision logic module that
dynamically allocates colors to cells. This module embodies the influential
primary factors that would operate on individual exit-choice decision
making within the context of descriptive behavior modeling. These
factors include the (i) group size of pedestrians along a path, (ii)
congestion at exits, (iii) width of exit gates, (iv) distance, and
(v) personal attitude to maintaining previous decisions.

\citet{haghaniSimulatingDynamicsAdaptive2019}, authorities in evacuation
behavior modeling, report that most research has been focused on modeling
directional decision making, while the decision adaptation phenomenon
has been largely overlooked. For example, exit-choice decisions at
the beginning and end of evacuations may be completely different.
An explicit factor is needed to model the variation in personal attitude
towards exit-choice changing. Thus, we include the 'personal attitude'
factor to model uncertainty and adaptation in exit-choice decision
making and study its influence on evacuation performance. Another
factor that plays a leading role in our MLM model is 'congestion at
exits', which is known to have a strong influence on pedestrians'
evacuation behaviors \citep{Liu20091921}. Typically, this factor
models the number of pedestrians at exits. However, this approach
neglects the exit gates' evacuation dynamics, which is crucial to
optimize the available capacity and to improve safety. For this purpose,
a method is proposed to characterize exit gates' evacuation dynamics.
This method is based on obtaining the Pedestrian Fundamental Diagrams
(i.e., the relationship between pedestrian flow and density) \citep{hoogendoornMacroscopicFundamentalDiagram2017},
and then applying a curve fitting function to parameterize each exit
gate in terms of congestion levels. Using these congestion levels,
we can define a safety metric and use it in the MLM simulation-optimization
process. 

A simulation-optimization modeling framework has been developed to
determine the optimal configuration of the MLM and obtain a near-optimal
adaptive evacuation plan. This framework integrates a microscopic
pedestrian simulation based on the classical Social Force Model (SFM)
\citep{helbingSocialForceModel1995}. The simulation-optimization
process adopts a Tabu-Search algorithm (TS) \citep{fredgloverTabuSearch1997},
which iteratively searches for the near-optimal evacuation plan (optimal
configuration of the MLM). At the same time, the microscopic crowd
simulation guides the search by evaluating the evacuation time and
safety of the solutions generated by the TS algorithm. The proposed
system is tested in a simulated real scenario (Madrid Arena) under
different external pedestrian flow patterns that simulate complex
pedestrian interactions. The research also presents a comparison between
traditional nearest-gate evacuation strategies and an existing approach
based on Cartesian Genetic Programming (CGP) \citep{Zhong2016127}
that we adapted to include the input factors used in our MLM. 

The rest of the paper is organized as follows. Section \ref{sec:Related work}
surveys the works related to crowd evacuation modeling. Section \ref{sec:Simulation-model}
outlines our modeling framework, which includes the evacuation scenario,
a proposal of system architecture for CellEVAC, the MLM used to build
the decision logic module of CellEVAC, and the microscopic simulation
framework to perform the simulation-optimization processes. The modeling
of exit gates' evacuation dynamics and safety are described in Section
\ref{sec:Modeling-pedestrian-flows}. Section \ref{sec:Experiments}
presents the experimental evaluation, results, and discussion. The
last section provides concluding comments and possible research extensions.

\section{\label{sec:Related work}Related work}

In this section we offer a summary of the previous work in the context
of crowd behavior modeling in evacuation scenarios, simulation modeling
and evacuation wayfinding algorithms.

\subsection{Exit-choice behavior modeling}

It is widely accepted that understanding the influence factors on
pedestrian behavior is fundamental in the design of large-scale public
facilities and evacuation planning. To capture the influence of these
factors and understand pedestrian behavior avoiding the potentially
dangerous real simulations, researchers have to rely on, for instance,
simulated environments \citep{bodeHumanResponsesMultiple2014,bodeInformationUseHumans2015}
or decision-making models built upon revealed and stated preferences
\citep{haghaniRandomUtilityModels2014}. 

The exit-choice strategies play a main role in the behavioral dimension
of evacuations \citep{hoogendoornPedestrianRoutechoiceActivity2004,kinatederExitChoiceEmergency2018,chenElementaryStudentsEvacuation2018,duivesExitChoiceDecisions2012}.
The existing research has identified numerous factors that explain
exit-choice decision making including the pedestrian emotion, distance
to exits, visibility, size of the queue, cooperation, route length
and capacity, illumination, or route familiarity \citep{duivesExitChoiceDecisions2012,cuestaEvacuationModelingTrends2015,lovreglioStudyHerdingBehaviour2016,haghaniPedestrianCrowdTacticallevel2016,kinatederExitChoiceEmergency2018}.
Assumptions about how an individual decides on an exit route, have
a significant influence on the overall evacuation effectiveness \citep{duivesExitChoiceDecisions2012,haghaniRandomUtilityModels2014,zhouSimulationPedestrianEvacuation2019}.
Some models assume that the exit-choice is only based on shortest
distance optimization \citep{hoogendoornPedestrianRoutechoiceActivity2004,klupfelModelsCrowdMovement2005}.
\citep{kinatederExitChoiceEmergency2018} also found that exit-choice
was influenced by exit familiarity and neighbor behavior. Using a
virtual crowd evacuation environment, \citep{bodeHumanExitRoute2013}showed
that individuals preferred routes with which they were familiar only
when they were presented with motivational messages. Also, they found
an evident influence of age and gender on reaction times. \citep{guoRouteChoicePedestrian2012}
revealed several behavior patterns related to preference for a destination,
effect of capacity, interaction between pedestrians, following behavior,
and evacuation efficiency. The trade-offs associated with these interactions
\citep{duivesExitChoiceDecisions2012,augustijn-beckersInvestigatingEffectDifferent2010,shiDevelopingDatabaseEmergency2009}
are present especially connected to the exit-choice decision. In \citep{liaoRouteChoicePedestrians2017},
the authors studied how time-independent and time-dependent information
affects subsequent changes in route choices and builds a simulation
model to illustrate simple behavioral mechanisms are sufficient to
describe the evacuation dynamics. They found evidence that dynamic
route choice behavior was comparatively rare. 

A growing body of literature has investigated how to model different
exit-choice behaviors in crowd evacuation simulations. Various approaches
have been proposed to solve this issue using discrete choice models
\citep{pressDiscreteChoiceAnalysis1985,ben-akivaDiscreteChoiceMethods1999}.
These models have been frequently used to define safety measures based
on guidelines discussing variables such as the exit door locations,
or the maximum density of people \citep{ronchiModellingLargescaleEvacuation2016,gaoBuildingEvacuationTime2020}.
Research in this field has focused primarily on the Multinomial Logit
Model (MLM) \citep{ben-akivaDiscreteChoiceMethods1999}, mainly used
to model the likelihood of individual choices from a discrete set
of alternatives.

In \citep{duivesExitChoiceDecisions2012} the authors investigated
an MLM to evaluate different exit-choice strategies. Their results
suggest that group following behavior has a significant impact on
evacuation. \citep{haghaniRandomUtilityModels2014,haghaniAccommodatingTasteHeterogeneity2015,haghaniHumanExitChoice2016,haghaniPedestrianCrowdTacticallevel2016,haghaniFollowingCrowdAvoiding2017,haghaniSocialDynamicsEmergency2017,haghaniStatedRevealedExit2017,haghaniCrowdBehaviourMotion2018}
reported on different methods to estimate random-utility models of
pedestrian exit-choice, and investigate crowd choice behavior during
evacuations of built environments. They propose a mixed (random-coefficient)
nested logit framework in \citep{haghaniAccommodatingTasteHeterogeneity2015}
and investigate underlying behavioral differences between normal egress
and emergency evacuations in \citep{haghaniHumanExitChoice2016} using
error-component mixed logit model of discrete choice analysis. \citep{lovreglioMixedLogitModel2016}
investigate the effect of environmental and social factors on local
exit-choice. They use an online stated preference survey using non-immersive
virtual reality, and a mixed logit model is calibrated.. More recently
\citep{haghaniFollowingCrowdAvoiding2017,haghaniSocialDynamicsEmergency2017}
report on wayfinding decision experiments that simulated the escape
from multi-exit spaces, and conclude that the assumption of herd-like
behavior does not necessarily apply to all contexts of evacuations.

Above mentioned studies have mainly dealt with the development of
descriptive models of evacuation behavior that simulate the real movement
of crowds \citep{duivesStateoftheartCrowdMotion2013,ronchiModellingLargescaleEvacuation2016}.
These models allow studying the influence of different behavior strategies
on evacuation performance \citep{zhouSimulationPedestrianEvacuation2019}.
However, behavioral strategy optimization is neglected to great degrees
\citep{Berseth2015377,nohEfficientPartiallyDedicated2016,dingSimulationbasedOptimizationEmergency2017}.
Among the body of studies on optimization, the majority have investigated
the problem from a path-planning \citep{vermuytenReviewOptimisationModels2016,nohEfficientPartiallyDedicated2016}
or architectural perspective \citep{zhaoOptimalLayoutDesign2017}.
Exit-choice behavioral optimization for designing evacuation guidance
systems remains a major knowledge gap that we address in this paper.

Recently, \citep{haghaniSimulatingDynamicsAdaptive2019} quantitatively
investigate the importance of including a decision changing module
for modeling adaptive decision-making in exit choices. They propose
a two-layered model with an exit-choice changing module and an exit-choice
module. The exit-choice changing module is a simple binary logit formula
that decides if pedestrians change their chosen exits. This formula
depends on directional attributes and a prefixed parameter that calibrates
the inertia to exit-choice changing. In case of change, the exit-choice
module, which implements a classical MLM, chooses a new exit. Results
showed a substantial difference in enhancing the accuracy of the simulation
outputs. They conclude that an intermediate degree of decision changing
is the most beneficial strategy. 

We have also explicitly paid particular attention to the exit-choice
changing phenomenon by modeling the pedestrians' attitude towards
changing their previous decisions. Instead of using a two-layered
model \citep{haghaniSimulatingDynamicsAdaptive2019}, our approach
uses a single-layered MLM that embeds both the directional and exit-choice
changing decision making components. This approach shortens the number
of parameters of the model, simplifying the optimization process.
Moreover, exit-choice decision changing is modeled as a time-dependent
personal attribute. It allows us to adapt the inertia to exit-choice
changing during the evacuation process, whereas in \citep{haghaniSimulatingDynamicsAdaptive2019}
the inertia is an immutable parameter. For instance, this approach
may be useful to reflect that confusion level changes during the evacuation
process, and therefore the level of inertia to decision changing.

\subsection{Simulation modeling}

Much of the related work on crowd behavior and evacuations must rely
on detailed simulations. In their recent survey of algorithms and
systems for evacuation, \citep{biSurveyAlgorithmsSystems2019} offer
state-of-the-art knowledge on emergency evacuation and \foreignlanguage{american}{wayfinding}.
According to their work, crowd behavior simulation models for evacuation
wayfinding can be classified into cellular automata models \citep{pelechanoEvacuationSimulationModels2008,felicianiImprovedCellularAutomata2016},
social force models \citep{helbingSocialForceModel1995}, fluid-dynamics
models \citep{hendersonStatisticsCrowdFluids1971}, lattice gas models
\citep{takimotoSpatiotemporalDistributionEscape2003}, game theoretic
models \citep{hoogendoornSimulationPedestrianFlows2003} and computer
agent-based models \citep{panMultiagentBasedFramework2007}. 

Many studies have been published using cellular automata models, which
discretize a given space into uniform ``cells'' where each cell
holds a person or vehicle \citep{DBLP:journals/access/Cruz-PirisML19}.
However, as mentioned by \citep{biSurveyAlgorithmsSystems2019}, these
models are ineffective at depicting movement speed and direction,
making it relatively challenging to customize physical attributes
or heterogeneous individuals with different characteristics. Social
force models (SFM) overcome these issues. These argue that pedestrians'
motion is mainly affected by the destination, the repulsive forces
from pedestrians and obstacles, and the attractive forces from other
objects (e.g., signals). Fluid-dynamics and lattice gas model-based
algorithms are better at simulating the movement of large crowds under
normal situations. Because of these approaches' macroscopic nature,
the modeling of complex interactions and individual behaviors such
as panic or uncertainty is difficult. Game-theoretic approaches \citep{8563}
model the cooperative and competitive behaviors during the evacuation
process, which is useful for mimicking the interactive decision-making
and strategy-adapting among evacuees. However, these approaches have
difficulty in capturing the dynamics of an evacuation process. Hence,
the agent-based models representing an environment with autonomous
decision-making agents, have drawn considerable attention in recent
years \citep{lopez-carmonaCooperativeFrameworkMediated2017}. Agents
have the ability to evolve, learn, and interact, which can lead to
unanticipated behaviors during simulations.

In our research, we opted for a multi-agent microscopic simulation
framework based on SFM due to its flexibility and ease of integration
of complex behavior and interaction models. This approach integrates
the potential of SFM to mimic physical interactions among evacuees,
and of multi-agent systems to simulate complex behaviors and interactions,
learn and evolve.
\begin{quotation}
\end{quotation}

\subsection{Evacuation wayfinding algorithms}

Many algorithms have been proposed \citep{biSurveyAlgorithmsSystems2019}
for the development of evacuation wayfinding systems. Network flow-based
algorithms consider evacuation planning as a minimum cost network
flow problem \citep{10.2307/j.ctt183q0b4}. The main downside of network
flow-based algorithms is that evacuees must follow the paths accurately
and reach every node on schedule. Various approaches have been put
forward to solve this issue using geometric graphs \citep{liLocalizedDelaunayTriangulation2003}.
For instance, in \citep{chenDistributedAreaBasedGuiding2008} a wireless
sensor network is partitioned into triangular areas based on the average
detected temperature of the associated sensors and safe egress paths
are calculated. A change of the topology induces redeployment and
re-calibration. More recently, \citep{guoPotentialbasedDynamicPedestrian2018}
proposed a potential-based dynamic pedestrian flow assignment model
to optimize evacuation time. Optimization is performed by employing
a space potential formulation and solving a proportional swapping
process between cells. Though the model is potentially applicable
during evacuation time, this work focuses only on obtaining an assignment
of pedestrians' exit choices at the beginning of the evacuation. Besides,
they propose two classes of guide sign systems that can guide pedestrians
to the corresponding exit.

Following this trend, queuing models \citep{newell2013applications}
transfer building graphs to a queuing network to estimate congestion
and evacuation delays. In \citep{chalmetNetworkModelsBuilding1982,yuhaskiModelingCirculationSystems1989,macgregorsmithStatedependentQueueingModels1991,pengwangModelingOptimizationCrowd2008,linoTuningValidationDiscreteEvent2011}
we may find studies in this area which are mainly focused on predicting
and optimizing the probabilistic choices in evacuations. Various approaches
dynamically develop navigation paths by assigning artificial potential
fields to the exits and hazards \citep{Koditschek,hillSystemArchitectureDirections2000,liDistributedAlgorithmsGuiding2003}.
Unfortunately, this mechanism suffers from several pitfalls, among
which is the convergence time for network stabilization, and the fact
that multiple exits may affect its search efficiency. 

There is a vast amount of literature on biologically-inspired algorithms
that employ heuristics to search for optimal routes or recommend exits.
In \citep{liMultiobjectiveEvacuationRoute2010} a multiobjective evacuation
route assignment model based on genetic algorithm \citep{hollandAdaptationNaturalArtificial1992a,gelenbeGeneticAlgorithmsRoute2006}
is proposed. This approach resembles the well known dynamic traffic
assignment (DTA) problem from the field of transport modeling \citep{bazzanIntroductionIntelligentSystems2013}.
In a similar approach, \citep{Abdelghany20141105} employed a simulation-optimization
framework that integrates a genetic algorithm and a microscopic pedestrian
simulation-assignment model. Evacuees are assumed to receive exit-choice
indications that may include the optimal start time of evacuation.
Its major drawback is that the calculated plan does not allow an adaptive
response. In \citep{Yadegari} bee colony optimization is used to
displace evacuees to safe areas. Its main drawback is the relatively
high communication overhead. \citep{Ferscha201033} developed a wearable
device named LifeBelt that recommends exits to individuals based on
the sensed environment. 

Similarly, in \citep{Zhong20141125} the idea is to use a gene expression
programming to find a heuristic rule. This rule is used to indicate
people in the same sub-region to move towards the same exit. The main
drawback of this solution is that it does not consider the dynamic
environment features in the evacuation planning process. Later in
\citep{Zhong2016127}, they propose a heuristic rule that considers
the distance and width of exit doors as fixed input parameters and
density around a given subregion as a dynamic parameter. The crowd
evacuation planning problem is converted to finding the optimal heuristic
rule that minimizes the total evacuation time. To solve this problem,
the authors adopt the Cartesian Genetic Programming (CGP) \citep{Miller2011}.
As we will show, the learning process using CGP is complicated, and
the control actions derived generate abnormal behaviors. Moreover,
they ignore the safety performance indicators in the optimization
process. Finally, in \citep{wongOptimizedEvacuationRoute2017} a shortest
path algorithm is used to compute routes by iteratively partitioning
graph edges at critical division points. Routes are iteratively refined
offline until an optimal state is achieved. This approach assumes
that a crowd distribution is known in advance, not considering safety
or dynamic changes during evacuation.

As in \citep{Abdelghany20141105}, our work develops a simulation-optimization
modeling framework that searches for the optimal evacuation plan through
meta-heuristic optimization methodology. However, we obtain adaptive
evacuation plans capable of responding to changing environmental conditions.
The CGP based crowd evacuation planning developed by \citep{Zhong2016127}
adapts to changing conditions, but its formulation is complex and
challenging to interpret. As mentioned above, the optimization process
is complex and difficult to configure and does not consider safety.
Moreover, the obtained evacuation heuristics functions for exit-choice
selection generate unnatural pedestrian movements. In contrast, the
CellEVAC system is easier to configure and optimize, and experiments
suggest a much more natural behavior. Also, we include a safety metric
in the optimization process. As far as we know, there are no studies
on how to include a safety metric in the simulation-optimization processes
to obtain evacuation plans.

\section{\label{sec:Simulation-model}Simulation-optimization modeling framework}

\subsection{Evacuation scenario}

Our investigation focused on Madrid Arena, an indoor arena located
in Madrid (Spain). The pavilion was designed to host sports events,
commercial, cultural and leisure activities. It has three floors (access,
intermediate, and ground) and 30,000 m\texttwosuperior , with a maximum
capacity of 10,248 spectators for basketball. Its central court has
three retractable bleachers, allowing the surface to change depending
on the type of event. On November 1, 2012, a stampede at a Halloween
party resulted in the death of five girls by crushing \citep{bbcnewsSpainHalloweenStampede2012}.
According to the police investigation, the cause was the excess of
capacity and the following errors in the indications of private guards
and police. There were not any guidance system to help evacuees choose
a safe exit gate.

\begin{figure}
\begin{centering}
\includegraphics[width=0.9\columnwidth]{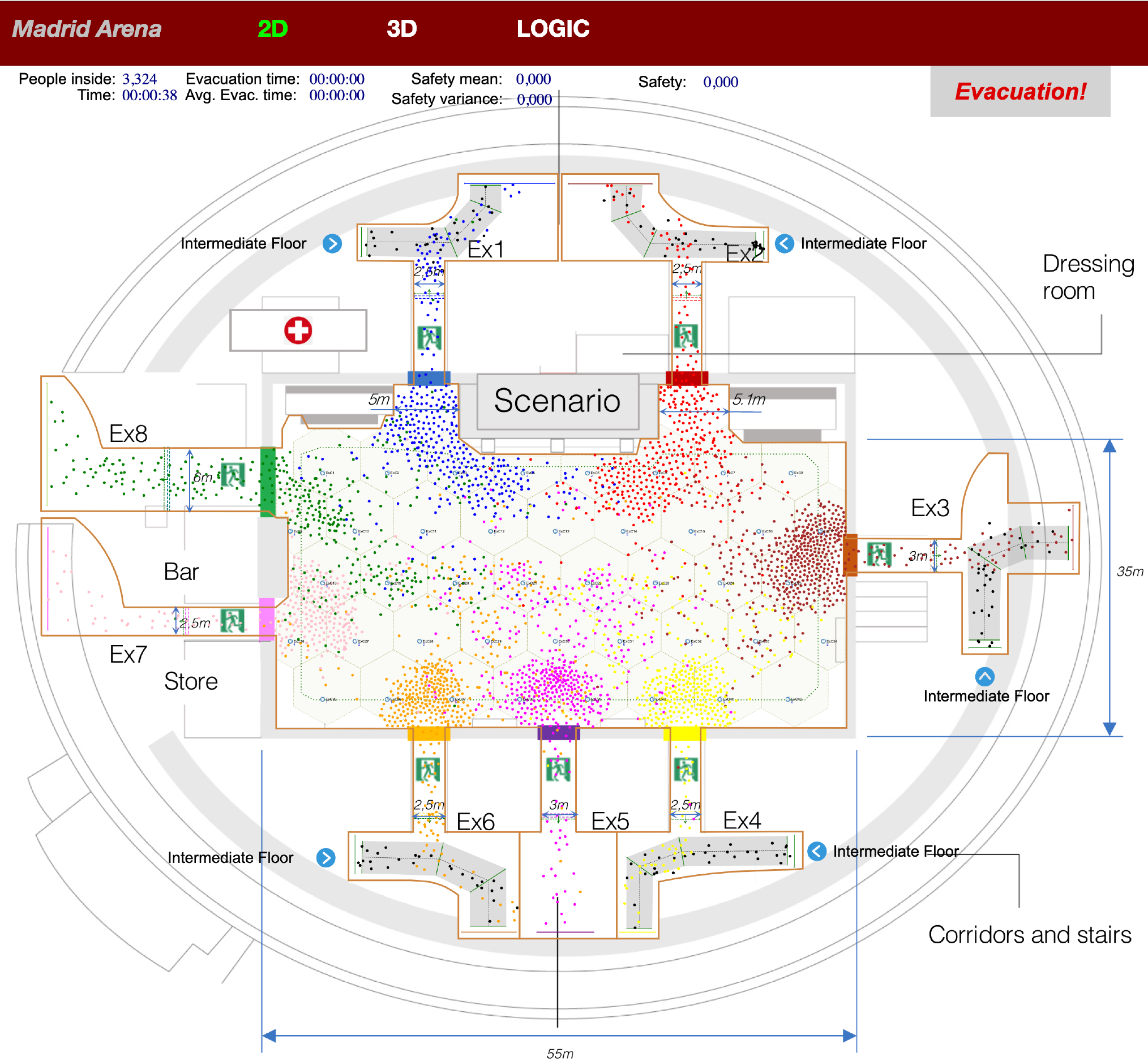}
\par\end{centering}

\caption{\label{fig:Madrid-Arena-layout.}Madrid Arena layout (ground floor).}
\end{figure}

We studied the evacuation of the ground floor, which has a maximum
capacity of $3,400$ spectators with the retractable bleachers removed.
Figure \ref{fig:Madrid-Arena-layout.} 
 shows the ground floor layout, with $1,925m^{2}$ and eight exit
gates (Ex1 to Ex8) with different widths between $2.5m$ and $6m$.
Each exit gate merges with exit corridors from the intermediate floors,
generating complex interactions between different pedestrian flows.
Pedestrian flows from intermediate floors were simulated by injecting
pedestrians at exits 1, 2, 3, 4, and 6 at the entry points highlighted
with a blue dot. 

\subsection{CellEVAC system architecture}

A system architecture was defined to illustrate a realistic deployment
of our CellEVAC guidance system in Madrid Arena. The architecture
assumes the utilization of existing off-the-shelf technologies, including
wearable devices or smartphones. The use of specific technologies
and its in-depth analysis and evaluation is beyond the scope of this
paper and is already underway. We aimed to conceive a generic architecture
focusing on the functional aspects of the system rather than on the
intrinsic elements of a specific technology. Although we include examples
of concrete technologies for some of the components of the architecture,
these are not claimed to be necessarily the only or best alternatives.
However, we believe that this proposal is generic enough to leave
room for exploring different technological alternatives.

\begin{figure}
\begin{centering}
\includegraphics[width=0.9\columnwidth]{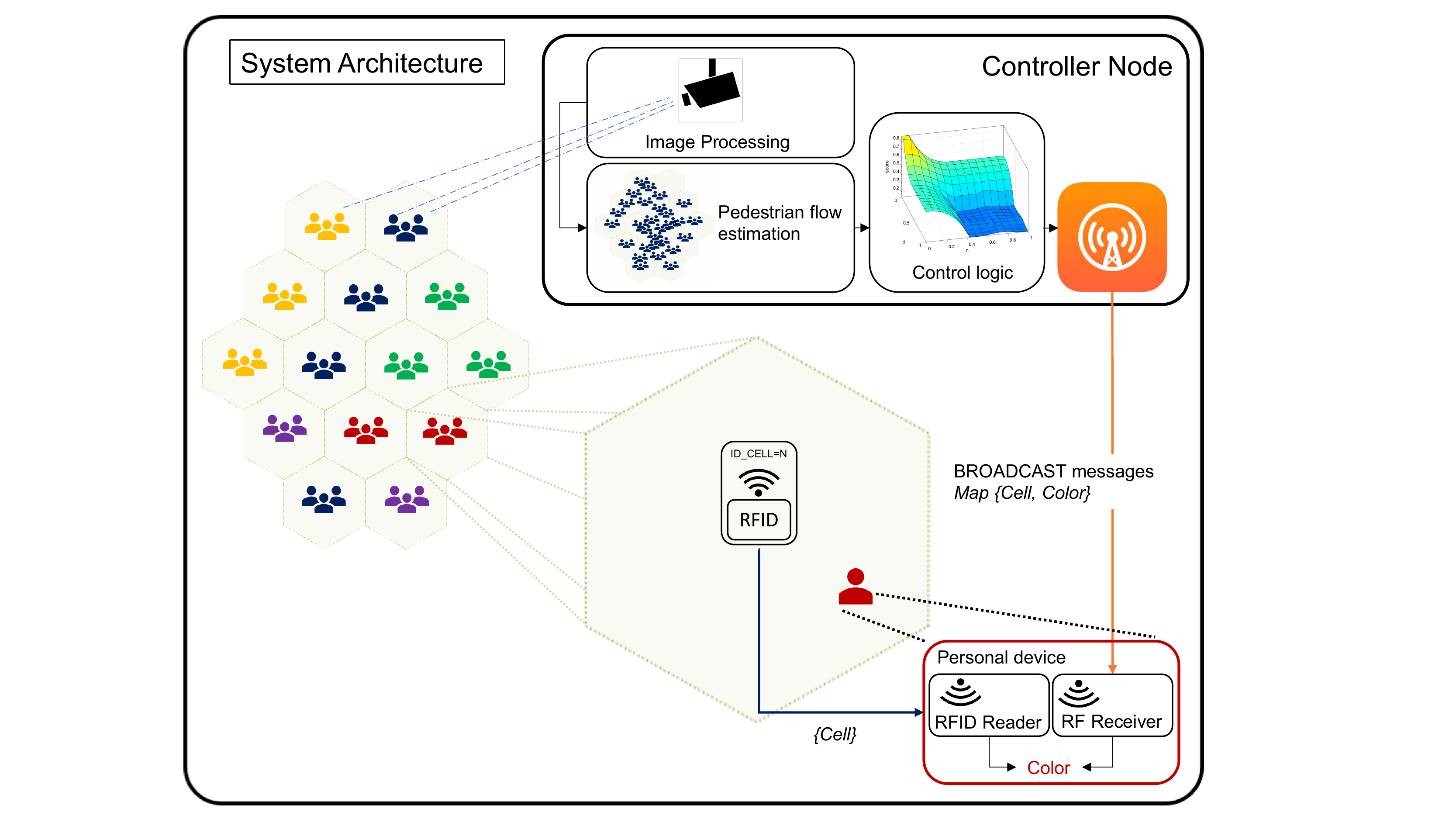}
\par\end{centering}
\caption{\label{fig:CellEVAC-System-Architecture}CellEVAC System Architecture.}
\end{figure}

For the specific case of Madrid Arena, the ground floor was divided
into 42 regular hexagonal cells of $9m^{2}$ and $6m$ width (Figure
\ref{fig:Madrid-Arena-layout.}). These dimensions were chosen to
provide a reasonable balance between control capability, wireless
coverage, and computational efficiency. Figure \ref{fig:CellEVAC-System-Architecture}
presents the system architecture, where a controller node embeds three
functional blocks: pedestrian flow estimation, control logic, and
Radio Frequency (RF) transmitter. The pedestrian flow estimation block
takes as input periodically sampled images that are preprocessed to
estimate pedestrian density at each cell. Obtained densities feed
the decision logic block to compute the optimal allocation of colors
to cells. With eight exits and 42 cells, we have a space of $2^{126}$
control actions at each control step. The RF transmitter broadcasts
messages periodically containing 42 tuples $\{Cell,Color\}$ that
assign a color to each cell. 

Each cell node is equipped with an active Radio Frequency Identification
(RFID) tag that periodically transmits a cell identification signal
to personal devices embedding an RFID Reader, whose purpose is to
provide location-aware capability. One real possibility to improve
cell detection accuracy is to implement a received signal strength
RFID-based indoor location mechanism \citep{alvarezlopezReceivedSignalStrength2017}.
The other module in the personal device is the RF Receiver. It periodically
receives the broadcast messages from the controller node. By matching
the pedestrian's cell-position and cell-color tuples, the personal
device lights up with the corresponding exit gate color.

From an implementation point of view, the most critical part of this
architecture is the positioning functionality. The RFID subsystem
that connects cells and personal devices have to cope with a complex
signal propagation environment and highly populated communication
channels. The RF channel is defined as a broadcast communication channel,
and consequently, it presents fewer problems. Finally, pedestrian
counting technology to estimate pedestrian density is widely available
in the market, though it should be tested and analyzed to assess its
adequacy depending on the specific application scenario.

\subsection{Pedestrian behavior modeling}

The first step in the development of the CellEVAC guidance system
was to model pedestrians' exit-choice decisions using a discrete choice
model. This choice was aimed at using the most widely applied theoretical
framework to model behavior in crowd evacuations. However, in contrast
to most existing research, our aim was not to calibrate the model
with real or survey data but to study the critical factors in exit-choice
decision making and then explore optimal pedestrian behavior strategies.
Furthermore, as mentioned in Section \ref{sec:Related work}, in an
attempt to simplify the model optimization, it was decided to use
a single-layered discrete choice model avoiding a separate structure
for exit-choice changing and exit-choice behaviors. For those readers
interested in an example of calibrated exit-choice decision model,
\citep{duivesExitChoiceDecisions2012} reported a comprehensive model
calibrated using responses to an Internet questionnaire conducted
in the Netherlands and United States. 

The most common theoretical framework for generating discrete choice
models is random utility theory \citep{antoniniDiscreteChoiceModels2006,ben-akivaDiscreteChoiceMethods1999,pressDiscreteChoiceAnalysis1985,ortuzarModellingTransport2011}
which postulates that: 
\begin{enumerate}
\item Individuals belong to a homogeneous population $P$, own perfect information
and act rationally.
\item There is a set $\mathbf{E}=\{E_{1},...,E_{j},...E_{N}\}$ of alternatives
and a set $\mathbf{X}$ of vectors of attributes of the individuals
and their alternatives. A given individual $p$ is provided with a
particular set of attributes $\mathbf{x}\in\mathbf{X}$ and will face
a choice set $\mathbf{E}(p)\in\mathbf{E}$.
\item Each option $E_{j}\in\mathbf{E}$ has associated a utility $U_{pj}$
for individual $p$. It is assumed that $U_{pj}$ can be represented
by two components:
\begin{enumerate}
\item a systematic part $V_{pj}$ which is a function of the measured attributes
$\mathbf{x}$; and
\item a random part $\varepsilon_{pj}$ which reflects the idiosyncrasies
of each individual and any observational error made. 
\end{enumerate}
Thus, we may postulate that:
\[
U_{pj}=V_{pj}+\varepsilon_{pj}
\]

$\mathbf{V}$ carries the subscript $p$ because it is a function
of the attributes $\mathbf{x}$ and this may vary from individual
to individual. It can be assumed that the residuals $\varepsilon$
are random variables with mean $0$ and a certain probability distribution
to be specified. A simple and popular expression for $\mathbf{V}$
is:
\[
V_{pj}=\sum_{k}\beta_{jk}x_{pjk}
\]

where the parameters (coefficients) $\beta$ may vary across alternatives
but are assumed to be constant for all individuals.
\item The individual $p$ selects the maximum-utility alternative $E_{j}$
if and only if:

\[
V_{pj}-V_{pi}\geqslant\varepsilon_{pi}-\varepsilon_{pj}
\]

Thus, the probability of choosing $E_{j}$ is:
\[
P_{pj}=\mathrm{Prob}\{\varepsilon_{pi}\leqslant\varepsilon_{pj}+(V_{pj}-V_{qi}),\forall E_{i}\in E(p)\}
\]

and as the joint distribution of the residuals is not known, different
model forms may be generated.
\end{enumerate}
In this research, we modeled exit-choice behavior using the simplest
and most popular practical discrete choice model, the Multinomial
Logit Model (MLM) \citep{duivesExitChoiceDecisions2012,ortuzarModellingTransport2011}.
The model is derived under the assumption that the error terms are
independent and identically distributed (IID) Gumbel (also called
Weibull or, more generally, Extreme Vale Type I). With this assumption,
the choice probabilities of exit $j$ by pedestrian $p$ are:

\begin{equation}
P_{pj}=\frac{\exp(V_{pj})}{\sum_{E_{i}\in E(p)}\exp(V_{pi})}\label{eq:Probabilistic}
\end{equation}
\\
The next step was to choose the attributes $x$ of the model (i.e.,
the factors that affect exit-choice decisions). The existing research
on exit-choice behavior has identified a broad range of influential
attributes \citep{haghaniSimulatingDynamicsAdaptive2019}: congestion
at exits, distance to exits, angular displacement, social influence,
visibility, exit width, and exit familiarity as well as personal characteristics
\citep{duivesExitChoiceDecisions2012,lovreglioDiscreteChoiceModel2014,kinatederExitChoiceEmergency2018}.

Several criteria were considered in choosing the attributes and structure
of the exit-choice model. The first criterion was the relevance of
the attributes informed by the literature and their relevance for
our specific evacuation scenario. The second criterion was considerations
about the number of attributes to include in the model. As stated
in \citep{haghaniSimulatingDynamicsAdaptive2019} for the specific
case of decision adaptation modeling, a long list of attributes poses
significant challenges to the problem of model calibration and optimization.
It imposes a limit on the number of attributes that the model can
reasonably include. The third consideration was the necessity of embedding
the directional and exit-choice changing decision making in a single-layered
modeling structure to simplify the optimization processes. Our approach
contrasts with the two-layered approach proposed in \citep{haghaniSimulatingDynamicsAdaptive2019},
which would roughly double the number of parameters to optimize. The
final consideration was the necessity of modeling the pedestrians'
response to indications given by evacuation guidance systems in general
and the CellEVAC system in particular.

After these considerations, we assumed that exit-choice decisions
were mainly motivated by the distance, congestion at exits, width
of exit gates, and the number of pedestrians along the path to each
exit gate (this is named 'group size of pedestrians'). It was decided
not to include visibility or angular displacement attributes because
the studied scenario had no obstacles. The primary purpose of differentiating
between congestion at exits and group size, was to model with more
precision the pedestrian flow dynamics in critical points. We aim
to model and optimize safety and capacity at exits using pedestrian
fundamental diagrams, and so, it makes sense to include in the guidance
systems a factor that evaluates congestion at the exits exclusively,
while group size measures congestion at inner areas. Note that both
the group size and congestion factors inherently model the tendency
to imitate behaviors. 

The current literature informs us that individuals tend to keep their
choices as much as possible \citep{liaoRouteChoicePedestrians2017}.
We reflected this in the model through a decision change attribute,
which captures the tendency to maintain the current exit-choice decision.
In \citep{haghaniSimulatingDynamicsAdaptive2019} this attribute is
included in the exit-choice changing model as a constant inertia parameter.
It implies that the tendency to change decisions is kept constant
during evacuation. Here, we chose a more general approach by making
the decision change factor time-dependent. It was reasonable to assume
that uncertainty in decision making evolves as evacuation progresses,
and so, the tendency to maintain the current decisions. Finally, we
reflected the influence of indications of the CellEVAC system through
a specific attribute that takes into account the exit-choice indications
given to each pedestrian.

Thus, the model for exit-choice is a Multinomial Logit Model based
on six attributes and as many alternatives as exit gates. The systematic
utility function for pedestrian $p$ and exit gate $j$ is given by: 

\begin{alignat}{1}
V_{\mathrm{\mathit{pj}}} & =\beta_{D}\times\frac{DISTANCE_{pj}}{max(DISTANCE)}+\beta_{W}\times\frac{WIDTH_{j}}{max(WIDTH)}\label{eq:Utility function}\\
 & +\beta_{G}\times\frac{GROUP_{pj}-GROUP_{min}}{GROUP_{pj}}+\beta_{E}\times\frac{EXCON_{j}}{criticalDensity_{j}}\nonumber \\
 & +\beta_{P}(t)\times PERSONAL_{pj}+\beta_{SYS}\times SYSTEM_{pj}\nonumber 
\end{alignat}
\\

The first attribute is the distance from pedestrian or cell center
$p$ to exit gate $j$, which is normalized in the range of 0-1 using
the maximum distance in the evacuation scenario, while the second
attribute represents the width of each exit gate, which is normalized
in the range of 0-1 using the maximum width.

The third attribute is the $GROUP$ ratio which estimates the congestion
level along a path from a pedestrian to an exit gate $j$, relative
to the congestion level at the least congested path. This estimate
is converted into a unitless attribute and confined within a fixed
interval 0-1, dividing it by the chosen path's congestion level. A
group ratio value of 0 would indicate that the chosen path is the
least congested path among the paths to the different exits. When
the value of the group ratio tends towards 1 for a given exit, it
means that the emptiest path's imbalance becomes large. Therefore,
the parameter $\beta_{G}$ is expected to be positive if pedestrians
tend to follow other pedestrians and is negative otherwise. Note that
with this type of normalization, the distribution of attribute values
does not exhibit a priori increasing or decreasing evolution over
time. Thus, we assumed that the relevance of the attribute in the
systematic utility function was kept constant throughout the evacuation
process.

The fourth attribute $EXCON$ accommodates the congestion at exit
gates. It was found reasonable to assume that pedestrians were able
to perceive both the density and flow of pedestrians to estimate the
congestion value. For a given density value, perceived congestion
is higher if the pedestrian flow is low. We reflect this effect through
critical density values obtained from the fundamental diagrams of
each exit gate (see Section \ref{sec:Modeling-pedestrian-flows}).
This $criticalDensity_{j}$ value reflects the density value at which
the exit gate's maximum capacity is reached. Therefore, the $EXCON_{j}$
value representing density at exit gate $j$ is normalized by the
corresponding $criticalDensity_{j}$ value. This normalization converts
$EXCON$ into a unitless attribute around 1. When the value of $EXCON$
is above 1, it means that exit is highly congested. A value close
to 0 would indicate that the exit gate is almost empty. In contrast
to the normalization procedure used for the $GROUP$ attribute, the
distribution of $EXCON$ values exhibits a decreasing evolution as
the number of pedestrians in the evacuation scenario decreases. It
seems reasonable to assume that the relevance of congestion at exits
as a discriminant factor for exit-choice decreases when the overall
number of pedestrians is low, and so $EXCON$ is close to 0 at all
exits. We recall that the $GROUP$ attribute's relevance is kept constant
during the evacuation process, which is in charge of capturing the
imitation behaviors, avoiding duplication of functions with $EXCON$.

The fifth attribute is the $PERSONAL$ value associated with person
$p$ and exit $j$, which captures how individuals are likely to revise
their previous exit-choice decision. We treat this attribute as a
binary categorical 0-1 value that equals 1 if the current exit-choice
of pedestrian $p$ is $j$, and is 0 otherwise ($PERSONAL=0\:\forall k\neq j$
). Therefore, in a general context, the parameter $\beta_{P}(t)$
is expected to be positive if pedestrians tend to maintain the previous
exit-choice, and is negative otherwise. However, we aimed to modulate
the tendency to maintain previous decisions, and so, $\beta_{P}(t)$
is always positive. As was noted above, we assumed that exit-choice
changing behavior evolves as evacuation progresses, and therefore
the parameter that modulates $PERSONAL$ is time-dependent. By observing
the pattern of behavior under various simulation settings, and considering
the optimization of the model, it was found reasonable that the tendency
to maintain decisions increased linearly depending on the current
number of pedestrians as follows:
\begin{equation}
\beta_{P}(t)=\beta_{P}\times\left(1-\frac{numOfPeds(t)}{numOfPeds(t=0)}\right)\label{eq:Beta-N}
\end{equation}

According to Equation \ref{eq:Beta-N}, at the beginning of the evacuation,
the parameter $\beta_{P}(t\rightarrow0)$ tends to $0$, and so, the
tendency to revise decisions is higher. As the number of pedestrians
to evacuate decreases, the parameter $\beta_{P}(t)$ tends to $\beta_{P}$,
and the tendency is to maintain previous decisions proportionally
to parameter $\beta_{P}$ value.

Finally, the $SYSTEM$ attribute measures pedestrians' attitude towards
the exit-choice indications of the CellEVAC system. We treat this
attribute as a binary categorical 0-1 variable such that $SYSTEM_{pj}$
equals 1 if pedestrian $p$ receives an indication to choose exit
$j$ (e.g., pedestrians see the color corresponding to exit $j$ in
their wearable devices), and is 0 otherwise.

It is important to emphasize that we do not claim that the set of
attributes is comprehensive, nor is the model using these attributes
claimed to be necessarily the fittest form. The model was kept as
simple as possible for two reasons. Firstly, to make model optimization
possible in a complex non-linear environment like this, it is important
not to face too many degrees of freedom. Otherwise, the search process
might be unmanageable. Secondly, to keep computational efficiency
to an acceptable level when simulating large crowds. Moreover, we
followed the recommendation stated in \citep{haghaniSimulatingDynamicsAdaptive2019}
to use unit-less attributes that help keep the model's generalizability
level. Note also that alternative specifications could potentially
be proposed even with the same attributes used in our model. For instance,
$EXCON$ relevance in the systematic utility function decreases during
the evacuation process, while $GROUP$ relevance is maintained constant.
Other alternatives could be possible depending on previous assumptions
about pedestrian behavior. 

Another important aspect of modeling exit-choice decisions that has
been reported in \citep{haghaniSimulatingDynamicsAdaptive2019} is
to mitigate decision changes that are not physically feasible. This
aspect mainly includes pedestrians confined by a crowded jam. In their
work, two measures are proposed to embody the decision-change eligibility
in the simulation model, based on measures of the local crowd density
and velocity experienced by pedestrians and on the definition of a
set of thresholds. This strategy could have been used in our model
by applying the same eligibility mechanisms and then apply or not
the probabilistic stage of the exit-choice process. However, we preferred
not to include this filter in our proposal for two reasons. Firstly,
to not increase the degrees of freedom and computational burden. Setting
the thresholds of eligibility to predefined values could bias the
exploration of optimal solutions. Secondly, for the specific case
of using the model to implement the decision logic of CellEVAC, our
experiments suggest that the movement is much more coordinated and
homogeneous. So the probability of having 'trapped' pedestrians is
lower.

Another aspect that impacts on the performance of the exit-choice
model, when applied to model pedestrian behavior or to implement the
decision logic of a guidance system, is the frequency at which the
pedestrians or the system revise their decisions. In the simulation
setting used in this work, we used an update cycle of 5 seconds. We
keep this frequency constant and control the frequency of the changes
at optimal levels using the explicit coefficient $\beta_{P}$ in the
model.

It is also important to note that the model allows us to establish
a clear distinction between environmental factors (distance, group,
width, and congestion), attitudinal factor (personal) and an exogenous
factor (system). Thus, the systematic utility function allows us to
model many different behaviors by trading off the different parameters.
For example, we may have a pedestrian who changes from making decisions
based on the indications received from CellEVAC to making decisions
based on environmental factors. However, it seems reasonable to assume
that a pedestrian following the indications is committed throughout
the entire evacuation process. Therefore, we focused on evacuation
scenarios in which a proportion of individuals (from 0\% to 100\%)
were committed to following indications, while the rest of the individuals
made their decisions based on environmental and attitudinal factors.
Given the extreme nature of emergency evacuations, we assumed that
individual characteristics tend to be reduced, and so all parameters
$\beta$ are defined as homogeneous values for each population group
throughout the evacuation time.

\subsection{Microscopic simulation-optimization framework}

A simulation-optimization software framework was developed embedding
agent-based simulation and discrete event simulation. The aim was
to simulate the CellEVAC system architecture integrating pedestrian
behavior modeling, SFM for pedestrian motion, control logic of exit
gate indications, and optimization features. We used as a basis the
commercially available programming, modeling and simulation software
packages AnyLogic \footnote{https://www.anylogic.com/ Accessed 19 June 2020}
and Matlab \footnote{https://www.mathworks.com/ Accessed 19 June 2020}.
The kernel of the simulation-optimization software framework is AnyLogic,
which provides multi-paradigm simulation, integrating three different
modeling methods: discrete event simulation, agent-based simulation,
and system dynamics, built on top of a Java-based software development
framework. We used extensively the AnyLogic pedestrian library, which
implements the Social Force Model (SFM) for simulating realistic pedestrian
motion \citep{helbingSocialForceModel1995}. Thus, with some particularities
described below, the evacuation scenario layout, pedestrian motion,
and evacuation measurements run in AnyLogic, while exit-choice decisions
and control logic of exit gate indications are implemented in Matlab.
AnyLogic and Matlab are interconnected in a master-slave configuration
through the interface with external Java libraries provided by AnyLogic
and the Matlab Java API engine (see details below). To study our CellEVAC
system based on MLM, and compare it with a Cartesian Genetic Programming
(CGP) approach, two simulation-optimization software environments
were developed: (a) simulation-optimization with MLM, and (b) simulation-optimization
with CGP, which are described in the following subsections.

It is worth noting that we do not mean that this combination of AnyLogic
and Matlab is the best possible simulation-optimization software framework.
Many different pedestrian evacuation simulators could be used individually
or combined with Matlab to build the models developed in this paper.
Our motivation to choose AnyLogic has been: the same development platform
can be used in different environments: supply chains, manufacturing,
transportation, rail logistics, warehouse operations; it provides
industry-specific libraries as pedestrian, rail, or road traffic libraries,
which eases the development work significantly; provides multimethod
simulation modeling: discrete event, agent-based, and system dynamics;
we can include discrete event processes, a multi-agent system, and
system dynamics in the same model; it is intuitive to build models
using graphical elements, leaving room to program at a very detailed
level; it provides gis maps integration, and it provides native capabilities
to develop simulations based on cellular automata or SFM. For the
interested reader, in \citep{lovreglioOnlineSurveyPedestrian2020},
an online survey of pedestrian evacuation model usage is presented
that studies the main trends in using pedestrian evacuation models
and simulators. 

\subsubsection{Simulation-optimization of MLM}

\begin{figure}
\begin{centering}
\includegraphics[width=1\columnwidth]{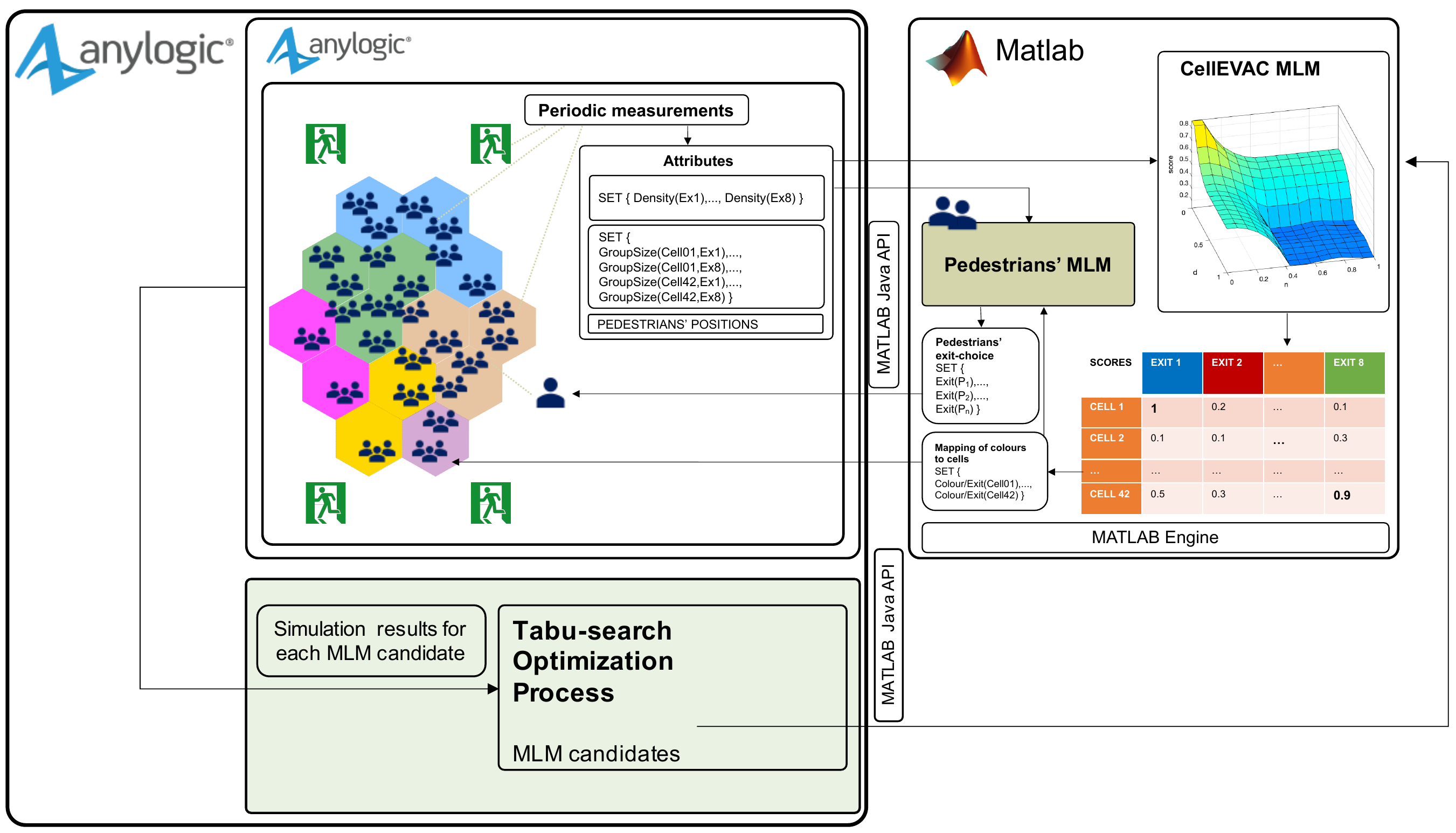}
\par\end{centering}
\caption{\label{fig:SimulationModelCellEVACMLM}Simulation-optimization software
framework of CellEVAC with control logic based on Multinomial Logit
Model (MLM).}
\end{figure}

The CellEVAC simulation model with MLM control logic is shown in Figure
\ref{fig:SimulationModelCellEVACMLM}. The evacuation scenario layout,
visualization features, and all the functionality regarding the SFM
based pedestrian motion were implemented within AnyLogic. 

During a simulation, the first step is to send from AnyLogic to Matlab
the set of parameters that configure the Pedestrians' MLM and CellEVAC
MLM modules, including the set of distances from the cell centers
to the exit gates. Next, the pedestrian positioning and densities
at exit gates and cells are periodically measured and then transformed
into the set of attributes: pedestrian positions, density at each
exit gate, and group of pedestrians along the path to each exit. To
obtain the number of pedestrians along a path, we used for convenience
the structure of cells defined for CellEVAC in Madrid Arena. For each
pedestrian within a given cell, group size is calculated by adding
the pedestrians in the cells that are closer to each exit. All these
attributes feed the CellEVAC MLM module in Matlab that implements
the decision logic to allocate colors (exit gates) to cells. This
mapping is sent back to AnyLogic for visualization purposes, and to
the Pedestrians' MLM module within Matlab to generate the individual
exit-choice decisions. 

Note that there is one MLM model (CellEVAC MLM) to generate exit gate
indications and a different MLM model (Pedestrians' MLM) used by pedestrians
to make exit-choice decisions. These decisions may follow the exit
gate indications through the $SYSTEM$ attribute of the Pedestrians'
MLM model. At one extreme, if we make $\beta_{SYS}=0$, pedestrians
will not follow exit gate indications from CellEVAC MLM. At the other
extreme, we could make all parameters $\beta=0$ except for $\beta_{SYS}=1$,
so that all pedestrians strictly follow the exit gate indications.
Also note that these two modules' systematic utility functions will
generally differ in the parameter values $\beta$ and in that the
CellEVAC MLM module does not include a $SYSTEM$ attribute. Moreover,
the $DISTANCE$ attribute in the CellEVAC MLM module corresponds to
the distance from a cell center to an exit gate, while in the Pedestrians'
MLM module it corresponds to the distance from a pedestrian to an
exit gate. 

Finally, while individual exit-choice decisions in the Pedestrians'
MLM model are probabilistic (see Equation \ref{eq:Probabilistic}),
the exit gate indications for a given cell in the CellEVAC MLM model
are based on a deterministic selection, corresponding to the exit
gate with the highest utility. The deterministic selection prevents
oscillations in the decision logic of CellEVAC.

As far as we know, this is the first time this simulation architecture
interconnecting AnyLogic and Matlab has been implemented. Although
AnyLogic is very valuable in the modeling of multiagent systems and
discrete event processes at a high level of abstraction, when developing
algorithms in very specialized domains (e.g., Control Systems, Fuzzy
Logic, Optimization, Deep Learning, ...) or when performing highly
intensive computational tasks (i.e., matrix computing), in our experience,
Matlab is a more productive computing environment. Fortunately, Matlab
provides flexible two-way integration with other programming languages.
With AnyLogic, we used the Matlab Engine API for Java, which enables
Java programs to use Matlab as a computational engine. AnyLogic imports
the Java Engine API library to enable two-way interactions with Matlab
synchronously or asynchronously. We found that this implementation
dramatically increased productivity in model development and experimentation,
facilitating the exploration of new decision making algorithms. For
example, evacuation simulations execute exit-choice decisions and
calculation of pedestrian distances to exit gates every few seconds
and for thousands of pedestrians. While in AnyLogic these tasks would
imply programming for-loops, Matlab built-in capabilities allow matrix
computation at a higher speed and with only a few code lines. 

To search for an optimal configuration of the CellEVAC MLM or pedestrians'
MLM models, we used a simulation-optimization process that adopts
a Tabu-Search algorithm (TS) \citep{fredgloverTabuSearch1997}, which
iteratively searches for the near-optimal evacuation plan (optimal
configuration of the MLM at the pedestrian level or system level).
At the same time, the microscopic crowd simulation guides the search
by evaluating the evacuation time and safety of the solutions generated
by the TS algorithm. The optimization process is built on top of the
OptQuest \footnote{https://www.opttek.com/ Accessed 22 June 2020}
optimization engine provided by AnyLogic. Figure \ref{fig:SimulationModelCellEVACMLM}
shows the optimization module on a green background. The parameters
of the CellEVAC MLM or pedestrians' MLM models are the ``MLM candidates''
generated by the TS algorithm. Thus, each candidate is defined by
a tentative set of parameters $\beta$ sent to the MATLAB Engine at
each iteration of the optimization process. The simulations results
are sent back to the optimization module for its evaluation and thus
guide the optimization process. 

\subsubsection{Simulation-optimization of CGP}

\begin{figure}
\begin{centering}
\includegraphics[width=1\columnwidth]{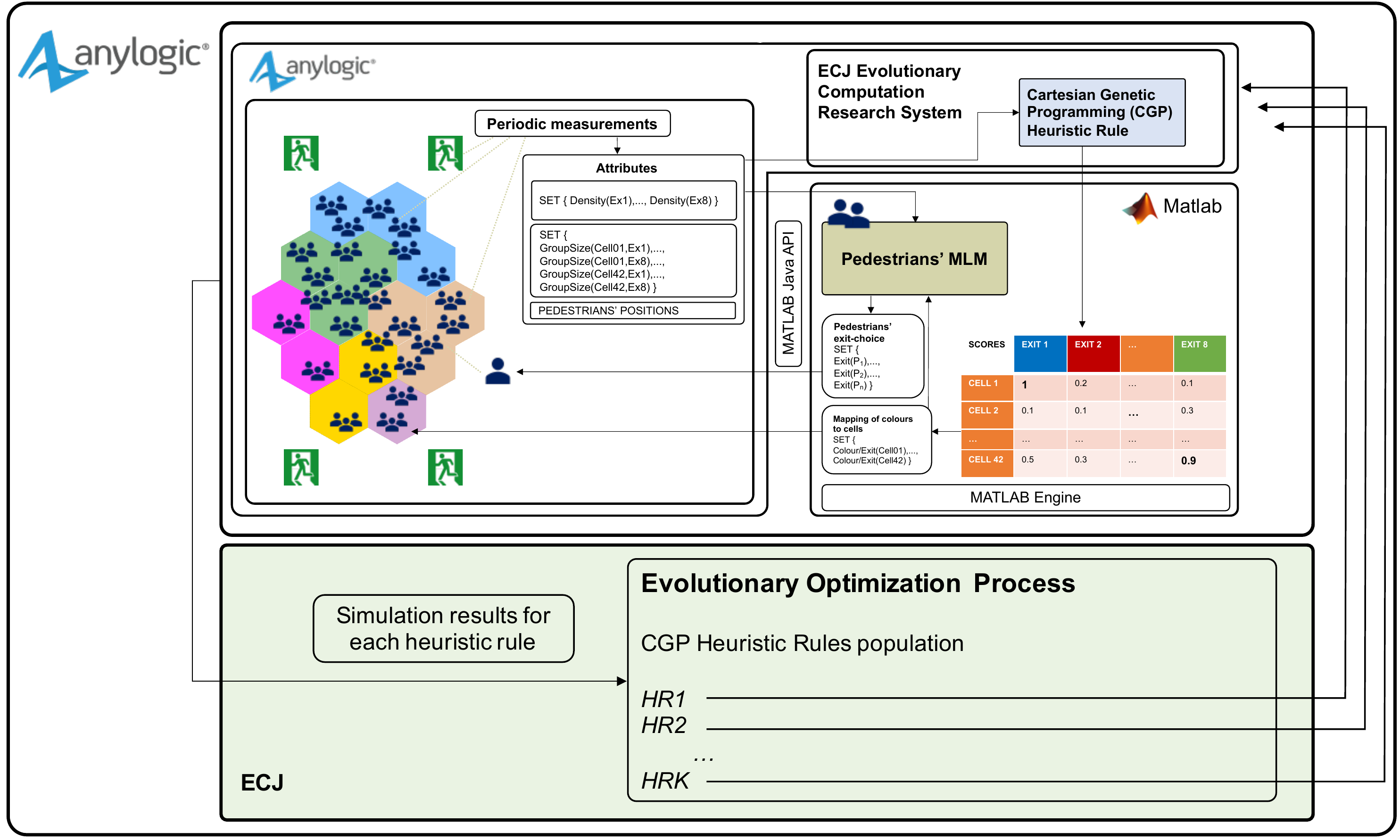}
\par\end{centering}
\caption{\label{fig:SimulationModelCellEVACCGPLearning}Simulation-optimization
software framework with control logic based on CGP.}
\end{figure}

Figure \ref{fig:SimulationModelCellEVACCGPLearning} illustrates the
simulation-optimization software framework that replaces the CellEVAC
MLM module with heuristic rules (programs) based on Cartesian Genetic
Programming (CGP) \citep{Miller2011}. A heuristic rule in CGP is
a program represented in the form of a directed acyclic graph as a
two-dimensional grid of computational nodes, which include input and
output nodes. In our application scenario, the input nodes receive
the attribute values of each pair cell-exit gate (i.e., distance from
cell center to exit gate, density at the exit gate, group size, width,
and tendency to maintain decisions), and a single output node returns
the score of each pair. As in the CellEVAC MLM model, selecting an
exit gate for a given cell is deterministic, corresponding to the
exit gate with the highest score.

Each node in CGP contains a set of integers that represents what operations
the node performs on the data and where a node gets its data. This
set of node values make up the genotype in the CGP. When the genotype
is decoded, some nodes may be excluded when they are not used to calculate
the output data. Thus, while the genotype in CGP has a fixed length,
the phenotype's size will be an intermediate value from 0 to the size
of the genotype. 

To obtain an optimal program, we used a variant on a simple evolutionary
algorithm known as a $1+\lambda$ evolutionary algorithm \citep{beyerEvolutionStrategiesComprehensive2002},
widely used for CGP. Although this algorithm could be implemented
using the Matlab Global Optimization Toolbox, the algorithms for decoding
or evaluating a CGP genotype must be implemented from scratch. It
was decided that the best procedure to implement the evaluation and
learning processes was to use the ECJ\footnote{https://cs.gmu.edu/\textasciitilde eclab/projects/ecj/ Accessed 3
March 2020} Java-based Evolutionary Computation Research System together with
the contribution package provided by David Oranchak\footnote{http://www.oranchak.com/cgp/doc/ Accessed 3 March 2020}
for CGP. ECJ is an evolutionary computation (EC) framework written
in Java. It provides tools that implement many popular EC algorithms
and conventions of EC algorithms but with a particular emphasis on
genetic programming. ECJ is free open-source with a BSD-style academic
license (AFL 3.0). 

We were able to integrate ECJ into AnyLogic and then use its full
potential as an open-source general-purpose evolutionary computation
framework. ECJ was imported as an external library within AnyLogic,
and some ECJ callback functions and AnyLogic functions were customized
to suit all simulation model requirements. As shown in Figure \ref{fig:SimulationModelCellEVACMLM},
the only difference compared to the CellEVAC framework is the processing
of the attributes to score the pairs cell-exit gate. In the CGP model,
attributes are processed by a CGP heuristic rule in Java within AnyLogic,
and the results are periodically sent to Matlab. To the best of our
knowledge, no other authors have developed this integration. 

To learn an optimal heuristic rule using CGP, we launch the ECJ evolutionary
process within a master AnyLogic experiment by calling an ECJ 'Evolve'
method (see the evolutionary optimization process module on a green
background in Figure \ref{fig:SimulationModelCellEVACCGPLearning}).
This method invokes an AnyLogic simulation experiment with CGP control
logic for each candidate heuristic rule within the population (see
the CGP Heuristic Rule module on blue background in Figure \ref{fig:SimulationModelCellEVACCGPLearning}).
The simulation results of each experiment are then sent back to the
evolutionary process in ECJ to evolve the population using a fitness
function (e.g., evacuation time and safety). This process is repeated
until the algorithm converges to an optimal heuristic rule.

\section{\label{sec:Modeling-pedestrian-flows}Modeling pedestrian flows and
safety}

It is well known that competition between agents at exit gates slow
down the evacuation processes. In the faster-is-slower effect described
by \citep{Helbing2000487}, waiting is seen to help coordinate activities
of competing agents and speed up the process, whereas more speed and
pressure slow the overall process. Increasing time-pressure causes
a severe reduction in the capacity of exits and a phase transition
from efficient free flow to congested flow. The main cause of capacity
drop for pedestrians is arc formation due to high pressure. Moreover,
when density becomes too large, the dynamics of flow are governed
by physical interactions generating hazardous situations. 

Evacuation systems should maintain density at exits under a critical
value to reduce this effect. Hence, to evaluate evacuation processes
and design effective evacuation systems, it should be deemed to characterize
flow dynamics at exits and consider both evacuation time and safety
metrics. To this end, the pedestrian Macroscopic Fundamental Diagram
(MFD) has proven to be a powerful concept in understanding and controlling
pedestrian flow dynamics. Similar to MFD for vehicular networks, a
relation exists between pedestrians' density and the average flow
(number of pedestrians per meter and per second) in an area. \citep{hoogendoornMacroscopicFundamentalDiagram2017}
provides results showing theoretically and empirically the existence
of pedestrian MFDs and the impact of density spatial variation on
MFDs. They conclude that the spatial density variation on flow is
likely to be dependent on the flow configuration. In general, decreasing
the spatial density variation will increase capacity. This fact has
implications in the construction of MFDs using microscopic simulation.
The shape of the MFDs will depend on the simulated flows' directions,
levels and time profiles.

In our proposal to derive an MFD for each exit gate, we used microscopic
simulation. Being our aim to obtain a safety metric using the fundamental
diagrams, we took a conservative approach. Each simulation entailed
pedestrians from four different directions heading to an exit to induce
a significative variation in spatial density and model complex interactions.
Thus, for each exit, four pedestrian flows were injected at four cells
around. Pedestrians had a preferred evacuation speed obtained from
a uniform distribution between 1.24 and 1.48 $m/s$. Each flow was
linearly increased for 10 minutes leading to exceeding capacity and
then linearly decreased to 0 for 10 minutes. This sequence was repeated
three more times to simulate queue build-up and recuperation phases
until a simulation interval of one hour was completed. At minute 50,
the exit was locked to characterize pedestrian dynamics in the event
of a fall. This repeating pattern and locking phase was based on experimentation,
aiming to generate shockwaves and a representative number of points
in the fundamental diagram.

Depending on the exit, peak flows ranged from 4 to 8 $peds/s$ to
exceed each particular critical density. With a two-second sampling
period, density was measured in an area defined by the four closest
cells to each exit while we took pedestrian flow measurements at each
exit gate. 

\begin{figure}
\begin{centering}
\includegraphics[width=0.75\columnwidth]{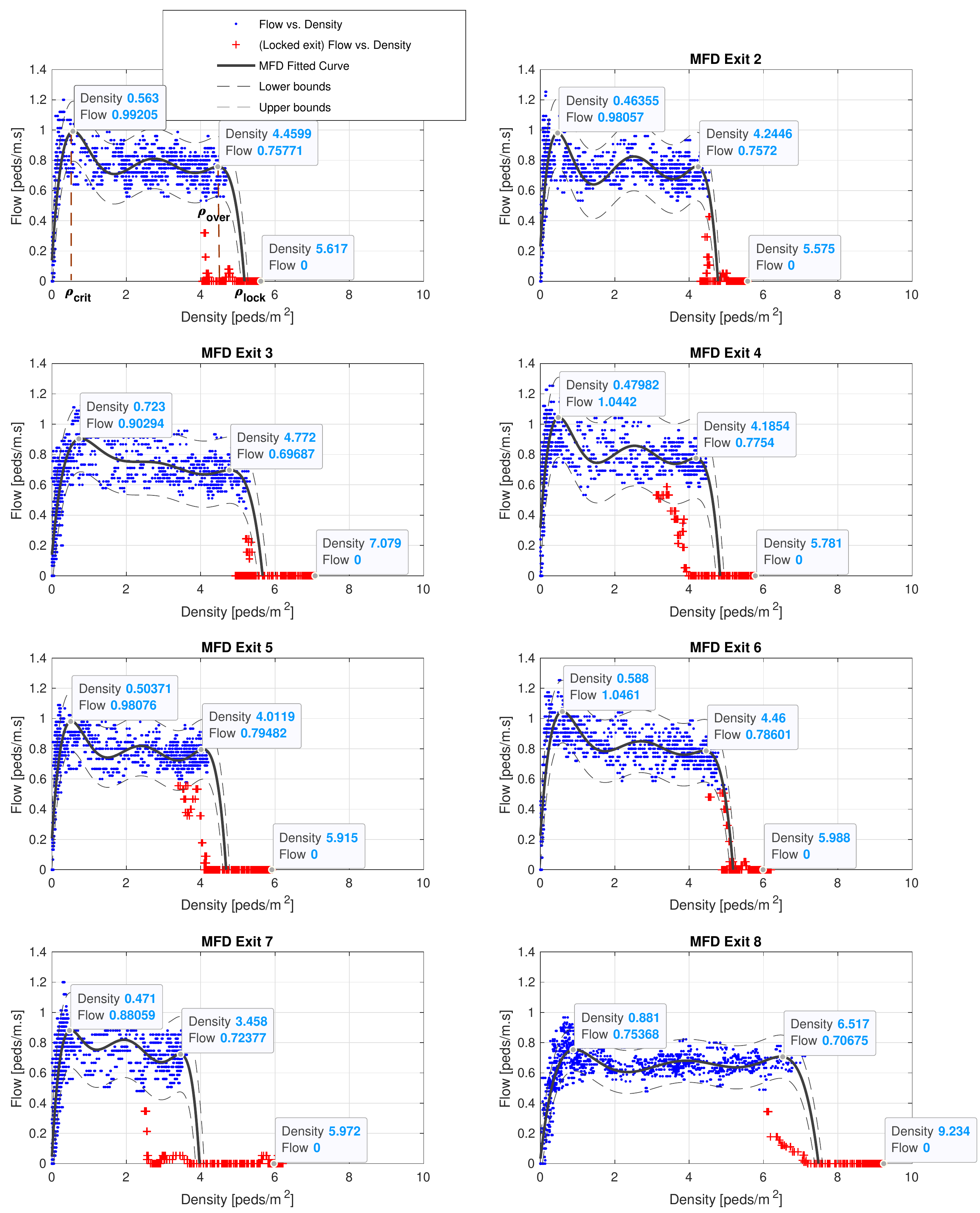}\caption{\label{fig:Fundamental-Diagrams-of}Fundamental Diagrams of exit gates.
The lower and upper bounds represent the 90\% confidence intervals.}
\par\end{centering}
\end{figure}

By following the procedure described above, we were able to obtain
the MFDs shown in Figure \ref{fig:Fundamental-Diagrams-of}. In blue
are represented the Flow vs. Density measurements during the first
50 minutes of simulation, while the red dots show the exit blocking
phase. We fitted curves to data using regression to characterize all
the points of interest in the MFDs homogeneously. We found the polynomial
model as the ideal candidate after exploring different alternatives
of curve fitting, such as smoothing splines, rational polynomials,
or gaussian models. Firstly, polynomials are often used when a simple
empirical model is required to obtain critical parameters. Furthermore,
we were able to easily parameterize the different phases of pedestrian
flow, looking at the fitted curve, and judging the reasonable values
for the critical densities. In all the cases, we used the bisquare
weights method for robust linear least-squares fitting and a sixth-order
polynomial.

In the MFD of Exit 1, it can be observed the different phases of pedestrian
flow during evacuation. Each phase is delimited by the density value
at which there is a flow peak. The critical density $\rho_{crit}$
delimits the free-flow region. During the simulations, we observed
that once the capacity value was reached, a fast backpropagation shockwave
was formed that rapidly carried the density value to $\rho_{over}$.
This state around $\rho_{over}$ is stable until flow decreases, and
a hysteresis path moves density to lower values. We observed that
this stable state maintained as long as the arc formation due to high
pressure was present. During the locking phase, the density value
increased beyond $\rho_{over}$ up to $\rho_{lock}$ due to queue
accumulation. When comparing the different MFDs, we found only slight
differences between the first seven exits, which is an expected result
considering that the evacuation scenario's geometry is quite regular,
without obstacles, and widths are similar. However, during simulations,
we observed a faster transition from $\rho_{crit}$ to $\rho_{over}$
at Exits 1 and 2 due to their funnel shape, which may contribute to
more dangerous situations at these exits. Exit 8 shows higher density
values but lower flow values. The only explanation for this is that
it is located in a corner. Note that lower flow values do not mean,
in this case, lower evacuation capacity. For instance, while at its
maximum capacity Exit 8 is able to evacuate 0.75{*}6 = 4.5 $peds/s$,
Exit 1 evacuates 2.5 $peds/s$. 

These density thresholds were used to build our safety metric for
evacuation. Firstly, the safety value at each exit gate $j$ is given
by the following equation:
\begin{equation}
Sf_{j}=(-\overline{\rho_{j}}-\gamma\cdot\sigma_{j}^{2})\times100,\label{eq:Safety}
\end{equation}
such that\\

\begin{align}
\overline{\rho_{j}} & =\frac{1}{N}\sum_{n=1}^{N}\frac{\rho_{j}'(n)-\rho_{sf_{j}}}{\rho_{lock_{i}}-\rho_{sf_{j}}}\:,\label{eq:mean-time density}\\
\sigma_{j}^{2} & =\frac{1}{N}\sum_{n=1}^{N}(\frac{\rho'(n)-\rho_{sf_{j}}}{\rho_{lock_{j}}-\rho_{sf_{j}}}-\overline{\rho_{j}})^{2}\:,\label{eq:variation-time density}
\end{align}
\\
with
\begin{align}
\rho_{j}'(n) & =\begin{cases}
\rho_{j}(n) & \rho_{j}(n)\geq\rho_{sf_{j}}\\
\rho_{sf_{j}} & \rho_{j}(n)<\rho_{sf_{j}}
\end{cases}\:,\label{eq:safety-function-normalization}\\
\rho_{sf_{j}} & =f(\rho_{crit_{j}},\rho_{over_{j}},\rho_{lock_{j}})\label{eq:weighted-average-function}
\end{align}

Given a set of $N$ density values measured periodically at exit gate
$j$, equations \ref{eq:mean-time density} and \ref{eq:variation-time density}
represent the time-mean density and density time-variation values
in the safety equation \ref{eq:Safety} respectively, which are weighted
by a configuration parameter $\gamma$. Density time-variation measures
the negative impact of variations of pedestrian flow, as explained
above. Both terms are normalized to 1 using the range defined by $\rho_{lock_{j}}$
and a predefined threshold $\rho_{sf_{j}}$. In Equation \ref{eq:safety-function-normalization},
those values below $\rho_{sf_{j}}$ are equal to $\rho_{sf_{j}}$
to make safety metric independent of the evacuation periods that are
considered safe. Thus, in a worst-case scenario in which exit gate
$j$ is locked, $Sf_{j}=-100$, while in a safe scenario where densities
are always below $\rho_{sf_{j}}$, $Sf_{j}$ equals 0. 

The value given to $\rho_{sf_{j}}$ (see Equation \ref{eq:weighted-average-function})
is defined as a function of $\rho_{crit_{j}},\rho_{over_{j}}$ and
$\rho_{lock_{j}}$. This function will depend on the specific scenario
and the availability of calibration data. Since in our study $Sf_{j}$
is used primarily for comparison purposes and not for its calibration
with real data, $f(.)$ was defined merely as the following weighted-average
function: 
\[
\rho_{sf_{j}}=0.9*\rho_{crit_{j}}+0.1*\rho_{over_{j}}
\]

Moreover, we decided to make $\gamma$ equal to 5 to strengthen the
influence of density time-variations.

Finally, to characterize the overall safety value of the evacuation
process, we used the average safety (Equation \ref{eq:Safety-Equation})
and variance of the safety (Equation \ref{eq:Safety_Variance}) values
at the exit gates:
\begin{align}
Sf= & \frac{1}{|E|}\sum_{j=1}^{|E|}Sf_{j}\:,\label{eq:Safety-Equation}\\
Sf_{var}= & \frac{1}{|E|}\sum_{j=1}^{|E|}(Sf_{j}-Sf)^{2}\label{eq:Safety_Variance}
\end{align}

The variance of safety $Sf_{var}$ was used to estimate the imbalance
of safety between the exit gates. Note that in contrast to the time-based
measurements used to calculate safety values at exit gates, safety
values to characterize the overall evacuation process are spatial-based
measurements.

\section{\label{sec:Experiments}Simulation-optimization experiments and results}

In this section, the experimental results are shown and discussed.
The performance measurements in all the experiments were the total
evacuation time, average and variance of safety values (Equations
\ref{eq:Safety-Equation} and \ref{eq:Safety_Variance}), and the
average number of decision changes. The users' average evacuation
time would be an alternative to total evacuation time, which could
be more suitable for evacuation scenarios where time to evacuate is
homogeneous, and there are no areas that need excessive time to evacuate.
However, our evacuation scenarios include external pedestrian flows
that increase the variance of evacuation times between exit gates. 

We conducted three main types of experiments: (i) sensitivity analysis
of exit-choice changing strategies, (ii) simulation-optimization of
pedestrians' behavior, CellEVAC and CGP systems, and (iii) performance
analysis of optimal configurations. In all the simulation setups,
the evacuee population consisted of 3400 pedestrians on the ground
floor, who had a preferred evacuation speed obtained from a uniform
distribution between 1.24 and 1.48 $m/s$. We considered an evacuation
scenario with external flows (i.e., with pedestrians coming from the
upper floors) to simulate complex pedestrian flow interactions. Therefore,
three exit gates were chosen at random at each simulation iteration.
Two of these exit gates received an incoming pedestrian flow rate
of $120\,peds/min$, while the third exit gate was blocked. We have
chosen 120 peds/min by visual inspection, increasing the external
pedestrian flows by hand until we found a flow at the corresponding
exits, creating significant congestion, close to the exit's maximum
capacity. 

In the sensitivity analysis experiments, each experiment ran the evacuation
simulation model multiple times varying one of the parameters, showing
how the simulation output (i.e., the performance measurements) depended
on it. Due to the stochastic nature of the evacuation processes, we
used a replication algorithm to obtain representative results for
a given parameter setting and a specific simulation output. This algorithm
defines a minimum and a maximum number of experimental runs per parameter
setting (replications of a simulation), a confidence level for the
sample mean of replications (simulation output average), and an error
percent. The minimum number guarantees the minimum number of replications,
while the confidence level and error percent determine if more replications
are needed. Simulation for a given parameter configuration stops when
the maximum number of replications has been run or when the confidence
level is within the given percentage of the mean of the replications
to date. In our experimental setup, evacuation time was used as an
output parameter to control the number of replications between 3 and
10. The confidence level was fixed to $80\%$, and the error percent
to $0.5$.

In the simulation-optimization experiments, the goal was to find the
combination of parameters of the MLM or CGP models that resulted in
the optimization of evacuation time and average safety through the
fitness function (objective function) $min(evacTime-Sf)$. We used
the Tabu-search optimization algorithm or the $1+\lambda$ genetic
algorithm depending on the model MLM or CGP to optimize. As in the
sensitivity analysis experiments, we used a replication algorithm
with the same basic setup. For each simulation-optimization process
iteration (i.e., for the model parameter setting to simulate and evaluate
at a given iteration), a minimum of 3 and a maximum of 10 simulations
were run for a total of 200 iterations. However, while in the sensitivity
analysis, the number of replicas was limited by the evacuation time
value, in simulation-optimizations, the stop condition was controlled
by the fitness function (objective function).

Finally, in the performance analysis experiments, the goal was to
evaluate the performance of a given model parameter setting. For instance,
for an optimal parameter setting found in a simulation-optimization
experiment, a performance analysis experiment ran the optimal setting
50 times for gaining statistical significance.

To analyze and compare the results of different experiments or configurations
we performed one-way analysis of variance \citep{montgomeryAppliedStatisticsProbablity1996a}
and Tukey-Kramer multiple comparison tests \citep{lauterHochbergAjitTamhane1989}.
For most of our experimental results, regardless of the type of performance
variables, the normality tests rejected the null hypothesis at 5\%
significance level. Thus, to provide a robust analysis of the results,
the Kruskal-Wallis test was used \citep{montgomeryAppliedStatisticsProbablity1996a}. 

\subsection{Sensitivity analysis of exit-choice changing strategies}

We first investigated how evacuation performance could be influenced
by exit-choice changing strategies in an evacuation scenario without
external flows, with the guidance system disabled (i.e., parameter
$\beta_{SYS}$ equals 0 in the pedestrians' MLM behavior model). 

\begin{figure}
\begin{centering}
\includegraphics[width=0.7\columnwidth]{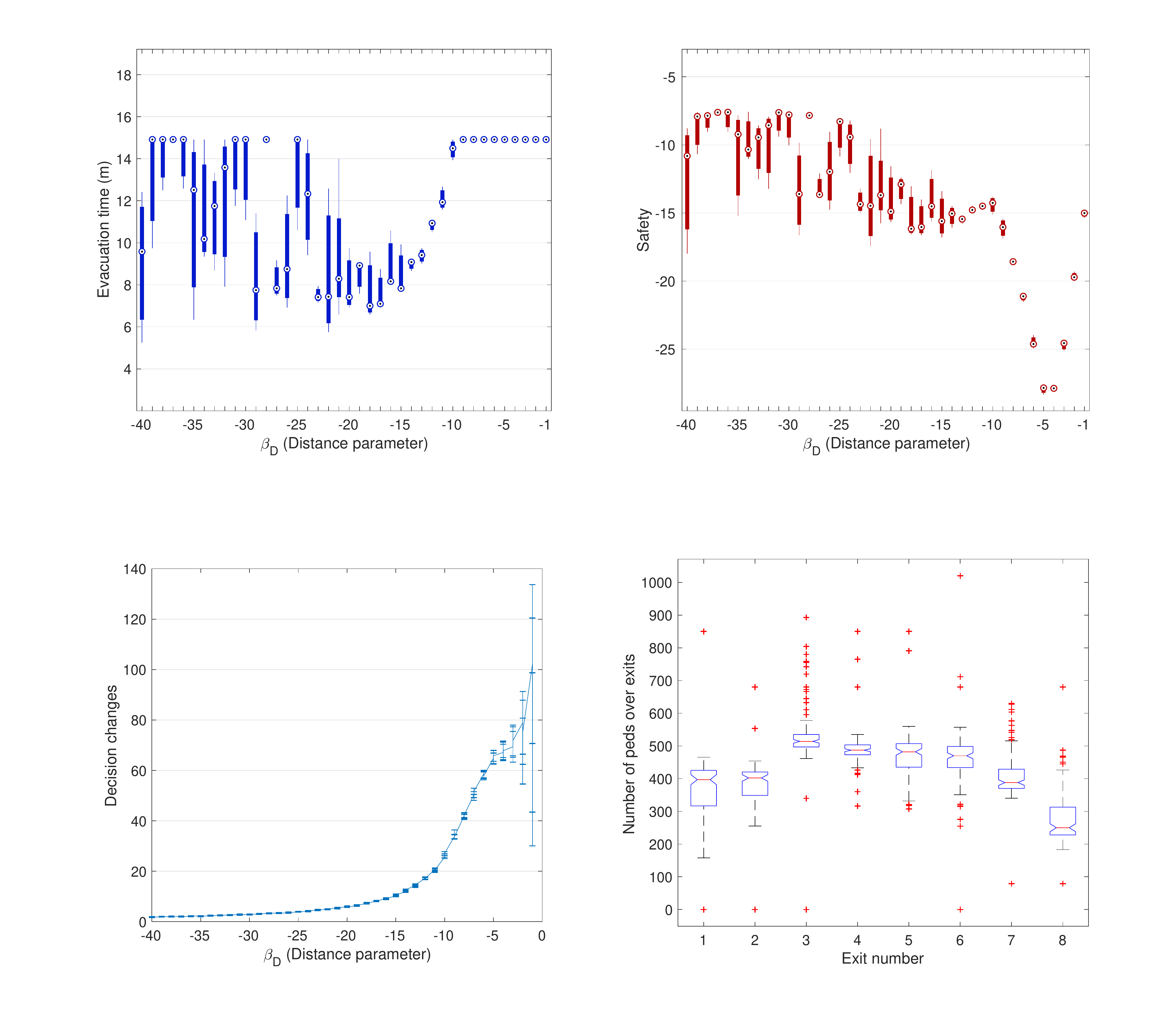}
\par\end{centering}
\caption{\label{fig:Sensitivity-analysis-of-B_DISTANCE}Sensitivity analysis
of the distance $\beta_{D}$ attribute without the guidance system
active. The box-plots on the first row show the sensitivity of evacuation
time and average safety to the $\beta_{D}$ attribute. The second-row
plots show the sensitivity of the number of decision changes to the
$\beta_{D}$ attribute, and the box-plots with the number of pedestrians
evacuated by each exit gate.}
\end{figure}

In these experiments, the only criterion for pedestrians' exit-choice
decision making was the distance. Parameter $\beta_{D}$ was varied
from $-40$ to $-1$ at discrete steps to perform a sensitivity analysis.
Evacuation time and safety results revealed pedestrians' irregular
behavior, with an unrealistic number of decision changes (see lower
box-plots in Figure \ref{fig:Sensitivity-analysis-of-B_DISTANCE}).
The visualization of simulations confirmed that the movement of pedestrians
was unnatural. As expected, as $\beta_{D}$ increased (i.e., the distance
criterion was less critical) uncertainty in decision making was higher,
the number of decision changes increased exponentially, and the safety
and evacuation time worsened (Figure \ref{fig:Sensitivity-analysis-of-B_DISTANCE}).
When $\beta_{D}$ was close to -1, pedestrians were not able to escape
from the evacuation scenario, and so the performance measurements
were distorted. Being the distance the only criterion for evacuation,
the number of pedestrians evacuated through the exit doors was unbalanced
(see box-plot of the number of peds over exits in Figure \ref{fig:Sensitivity-analysis-of-B_DISTANCE}).
For instance, exit gate 8 evacuated the least number of pedestrians,
though it is the exit with the highest capacity. Overall, these results
revealed the necessity of including an exit-choice changing attribute
in the MLM model to modulate the number of decision changes.

\begin{figure}
\begin{centering}
\includegraphics[width=0.65\columnwidth]{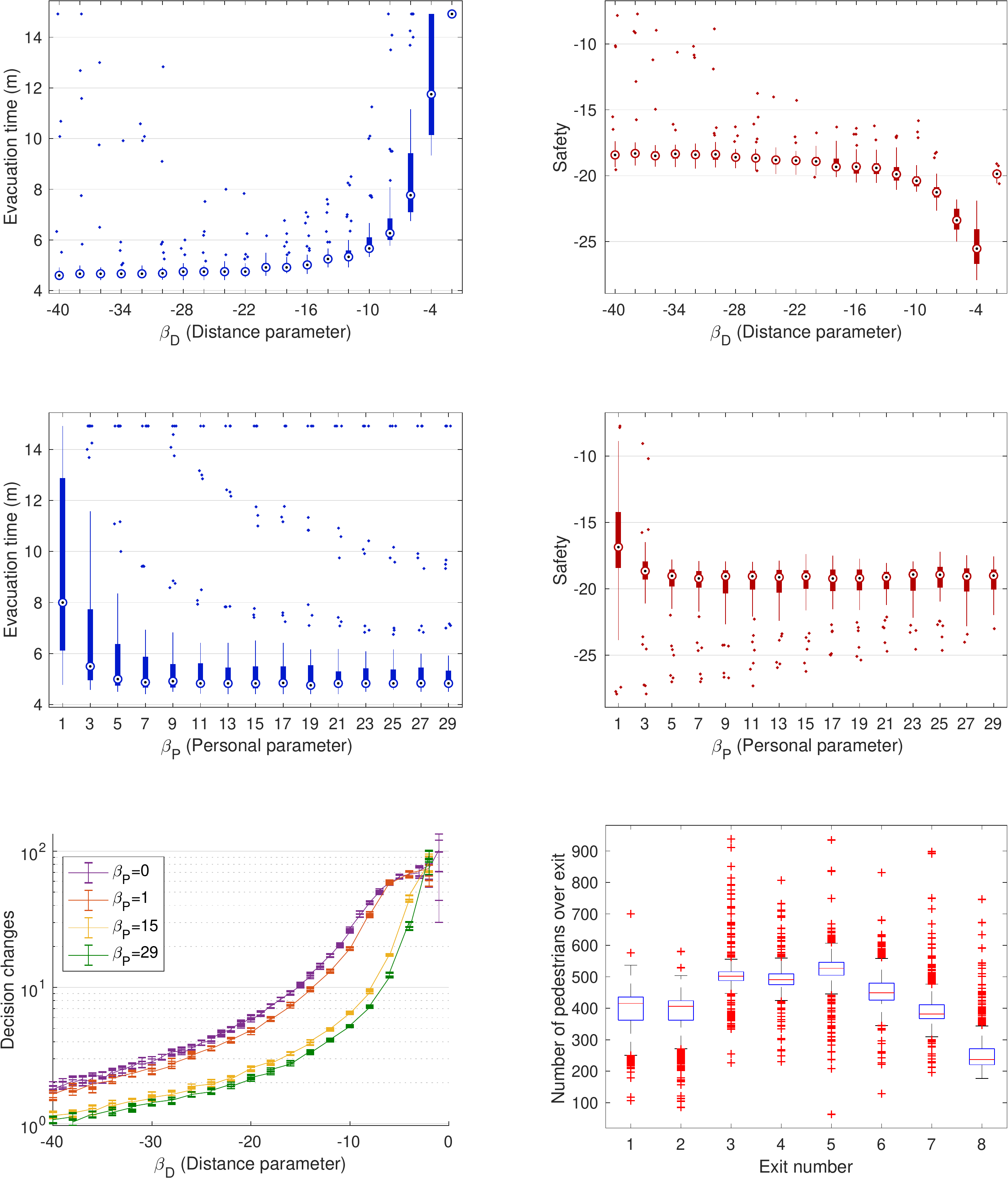}
\par\end{centering}
\caption{\label{fig:Sensitivity-analysis-of-B_DISTANCE-B_PERSONAL}Sensitivity
analysis of $\beta_{\mathrm{D}}$ and $\beta_{\mathrm{P}}$ without
external pedestrian flows and the guidance system inactive. The box-plots
on the first and second rows show the sensitivity of evacuation time
and average safety to the $\beta_{D}$ and $\beta_{P}$ attributes.
The plots on the third row show the sensitivity of the number of decision
changes to the $\beta_{D}$ attribute, and the number of pedestrians
evacuated by each exit gate.}
\end{figure}

The next experiments extended the previous sensitivity analysis to
include the $PERSONAL$ attribute, which models exit-choice changing
behavior. Parameter $\beta_{P}$ was varied in the range 1 to 29 at
discrete steps to modulate the tendency to maintain the current exit-choice
decision. Figure \ref{fig:Sensitivity-analysis-of-B_DISTANCE-B_PERSONAL}
presents the results of the sensitivity analysis. The first row shows
the results of the experiments, in which for a given $\beta_{D}$
value, $\beta_{P}$ is varied in the range 1 to 29 (i.e., for a given
$\beta_{D}$ value, we performed 29 evacuation experiments with a
different $\beta_{P}$ value). In the second row, the box-plots show
for a given $\beta_{P}$ the results of 40 evacuation experiments
with a different $\beta_{D}$ value. In the decision changes box-plot
(third row), for each $\beta_{D}$ value, we perform four experiments
with $\beta_{P}$ equal to 0, 1, 15 and 29. 

The results obtained confirmed the hypothesis of a much more stable
and realistic pedestrians' behavior. For a wide range of values of
both parameters $\beta_{D}$ and $\beta_{P}$, evacuation time and
safety values were confined within a small range of values. Moreover,
evacuation time significantly improved when compared to not using
the $PERSONAL$ attribute. For instance, for a value of $\beta_{D}=-22$,
median evacuation time was around $4.8$ minutes, while without the
$PERSONAL$ attribute, it varied in the range 6-12 minutes. 

In contrast, the safety value worsened from -14 to -19 with the $PERSONAL$
attribute. These results suggest that there is a correlation between
safety and the number of decision changes. Uncoordinated movement
of pedestrians far from the exit gates made the evacuation time increase,
density at exit gates decrease, and so safety at exits improve. As
regards the results for the number of decision changes, these revealed
a significant reduction in the number of decision changes as $\beta_{P}$
increased. When combined with the evacuation time and safety results,
it can be seen that the optimal strategy is to make very few exit-choice
changing decisions ($\beta_{P}=29$) based on a highly influential
distance criterion ($\beta_{D}=-40$). Finally, as in the previous
experiment, there was not a fair balance between the different outflows.
We replicated the experiments in an evacuation scenario with external
flows. The results obtained are similar to those outlined above, except
for the magnitude of the evacuation performance results.

These experiments are consistent with previous findings \citep{haghaniSimulatingDynamicsAdaptive2019}.
According to our results, it is evident that the inclusion of the
decision change $PERSONAL$ attribute made a very substantial difference
in terms of the efficiency of evacuations. Moreover, the experiments
confirmed the viability of including the modeling of exit-choice changing
behavior in a single-layered MLM structure.

\subsection{Simulation-optimization of pedestrians' behavior (OPTIMAL), the CellEVAC
and CGP systems}

Here we investigated the optimal individual behavior of pedestrians
(OPTIMAL), and the optimal configurations of the CellEVAC and CGP
guidance systems. In Appendix \ref{sec:Optimization-processes} are
described in detail the conducted optimization processes. 

The optimal settings found for OPTIMAL and CellEVAC have been summarized
in Table \ref{tab:Optimal-configuration-of-Pedestrians-MLM}. The
STANDARD parameter setting in Table \ref{tab:Optimal-configuration-of-Pedestrians-MLM}
models typical pedestrian behavior in which exit-choice decisions
are based mainly on distance, and to a lesser extent, on imitation
and visual perception of the exit gates' width. These values are based
on the parameter setting obtained in \citep{duivesExitChoiceDecisions2012}
for a calibrated model. However, these values are not claimed to be
necessarily the fittest form of standard behavior, but a typical behavior
strategy to compare. Finally, we found the following heuristic rule
for the CGP system:

\begin{multline*}
score(cell_{i\in\{1,...,42\}},exit-gate_{j=\{1,...,8\}})=\\
*(-(-(-WW)G)(-P(+G(-PD))))\\
(+(+(-(-(/(/P(-PD))(/(/P(-PD))\\
(-PD)))(-WW))\\
(-P(/P(+G(-PD)))))\\
(*(/(/P(-PD))\\
(/(/P(-PD))(-PD)))E))\\
(+(/(-PD)(+(/P(-PD))(-(-WW)G)))E))\\
\end{multline*}

where $D=DISTANCE$, $E=EXCON$, $G=GROUP$, $W=WIDTH$ and $P=PERSONAL$.

\begin{table}
\begin{centering}
\begin{tabular}{cccccc}
\hline 
 & {\small{}$\beta_{\mathrm{D}}$} & {\small{}$\beta_{\mathrm{G}}$} & {\small{}$\beta_{\mathrm{E}}$} & {\small{}$\beta_{\mathrm{W}}$} & {\small{}$\beta_{\mathrm{P}}$}\tabularnewline
\hline 
{\small{}STANDARD} & {\small{}-28} & {\small{}0.6} & {\small{}-0.5} & {\small{}0.6} & {\small{}0}\tabularnewline
{\small{}OPTIMAL} & {\small{}-28.863} & {\small{}9.909} & {\small{}-2.801} & {\small{}0} & {\small{}8.515}\tabularnewline
{\small{}CellEVAC} & {\small{}-17.723} & {\small{}-2.181} & {\small{}-1.671} & {\small{}1.064} & {\small{}2.594}\tabularnewline
\hline 
\end{tabular}
\par\end{centering}
\caption{\label{tab:Optimal-configuration-of-Pedestrians-MLM}Optimal and standard
configurations of the parameters of the pedestrians' MLM behavior
model. }
\end{table}

The obtained parameter values for OPTIMAL revealed that the most influential
factor was the distance. Interestingly, we found that group imitation
behavior ($\beta_{G}=9.909$) had a positive effect. Our interpretation,
corroborated by visual inspection, is that collective intelligence
contributes to a better balance in exit gate sharing in the presence
of complex pedestrian flow interactions.

The negative sign of $\beta_{E}$ (congestion at exit gate) highlighted
the benefit of avoiding congested exit gates. However, the exit gate's
width was found irrelevant in scenarios with external flows. These
results are in line with our hypothesis of a better evacuation performance
in complex environments when using adaptive strategies. Finally, the
value obtained for exit-choice changing parameter $\beta_{P}$ confirmed
the positive effect of gradually smoothing the number of decision
changes during the evacuation process.

For CellEVAC, the parameter values also revealed that the most influential
factor was the distance. Remarkably, we found that in contrast to
OPTIMAL (i.e., assuming that individuals behave optimally at an individual
level), the group parameter $\beta_{G}$ had a negative sign. Our
interpretation is that the group imitation effect is implicit in the
cell-based control of pedestrian movements, so a positive value of
this parameter negatively influences an excessive uniformity in the
exit gate indications. Therefore, the $\beta_{G}$ parameter compensates
for the implicit grouping effect of CellEVAC. We also observed significant
differences in magnitude in the exit gate width and decision changing
parameters. The tendency to maintain previous cell indications by
the CellEVAC system was significantly lower than to main previous
exit-choice decisions at the individual level (2.594 vs. 8.515). The
width of the exit gate, which was not considered at the individual
level, appeared at system level. 

The interpretation of the heuristic rule for the CGP system is not
straightforward, and we would need to apply partial derivatives to
extract information regarding the influence of the different attributes.
Also, due to the simulation processes' stochastic nature in CGP, each
optimization trial may obtain different heuristic rules, making generalization
difficult.

\subsection{Performance results}

\begin{figure}
\begin{centering}
\includegraphics[width=0.6\columnwidth]{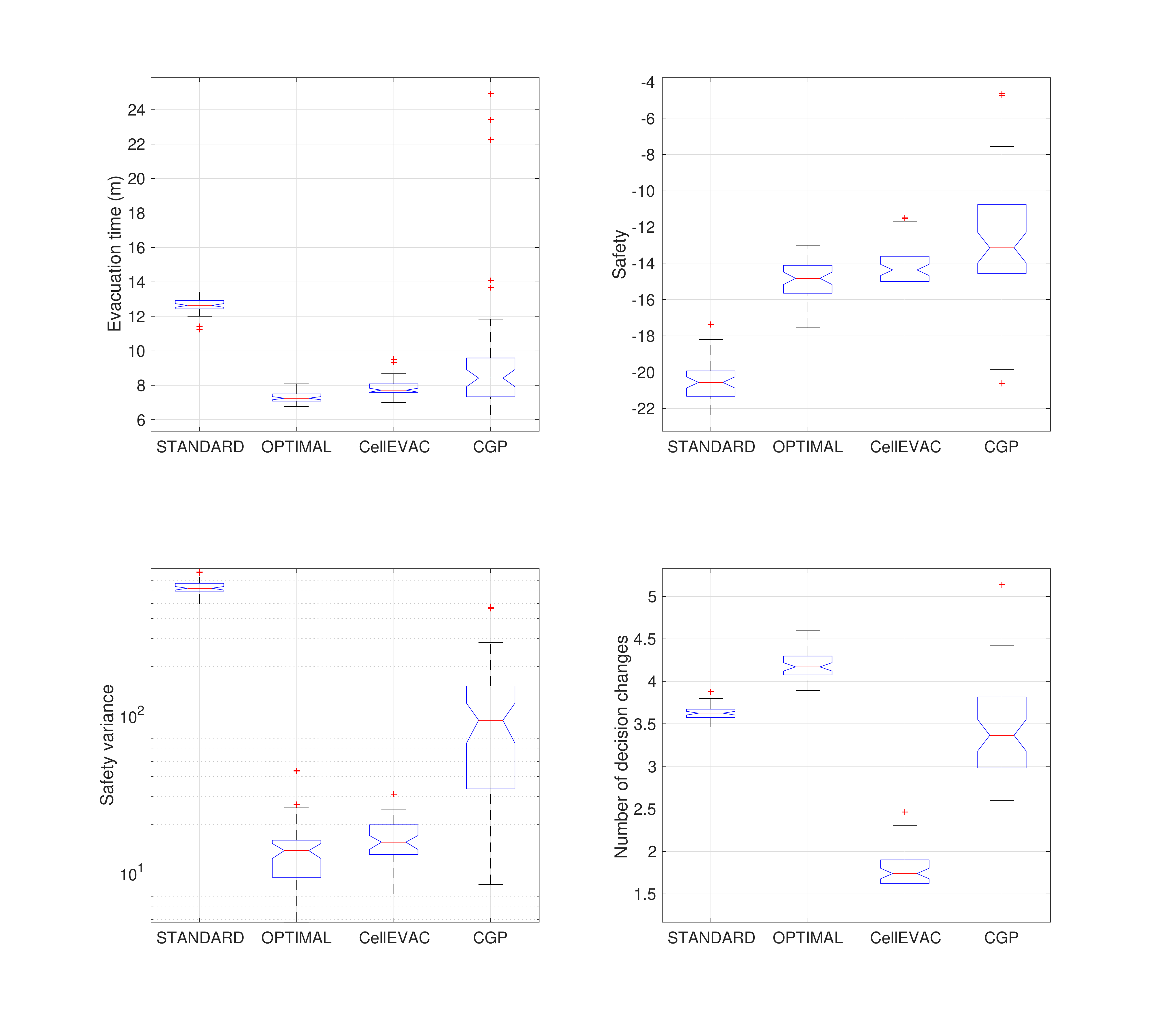}
\par\end{centering}
\caption{\label{fig:Boxplots-General}Box-plots of the performance measurements
for the STANDARD, OPTIMAL, CellEVAC and CGP configurations.}
\end{figure}

\begin{figure}
\begin{centering}
\includegraphics[width=0.5\columnwidth]{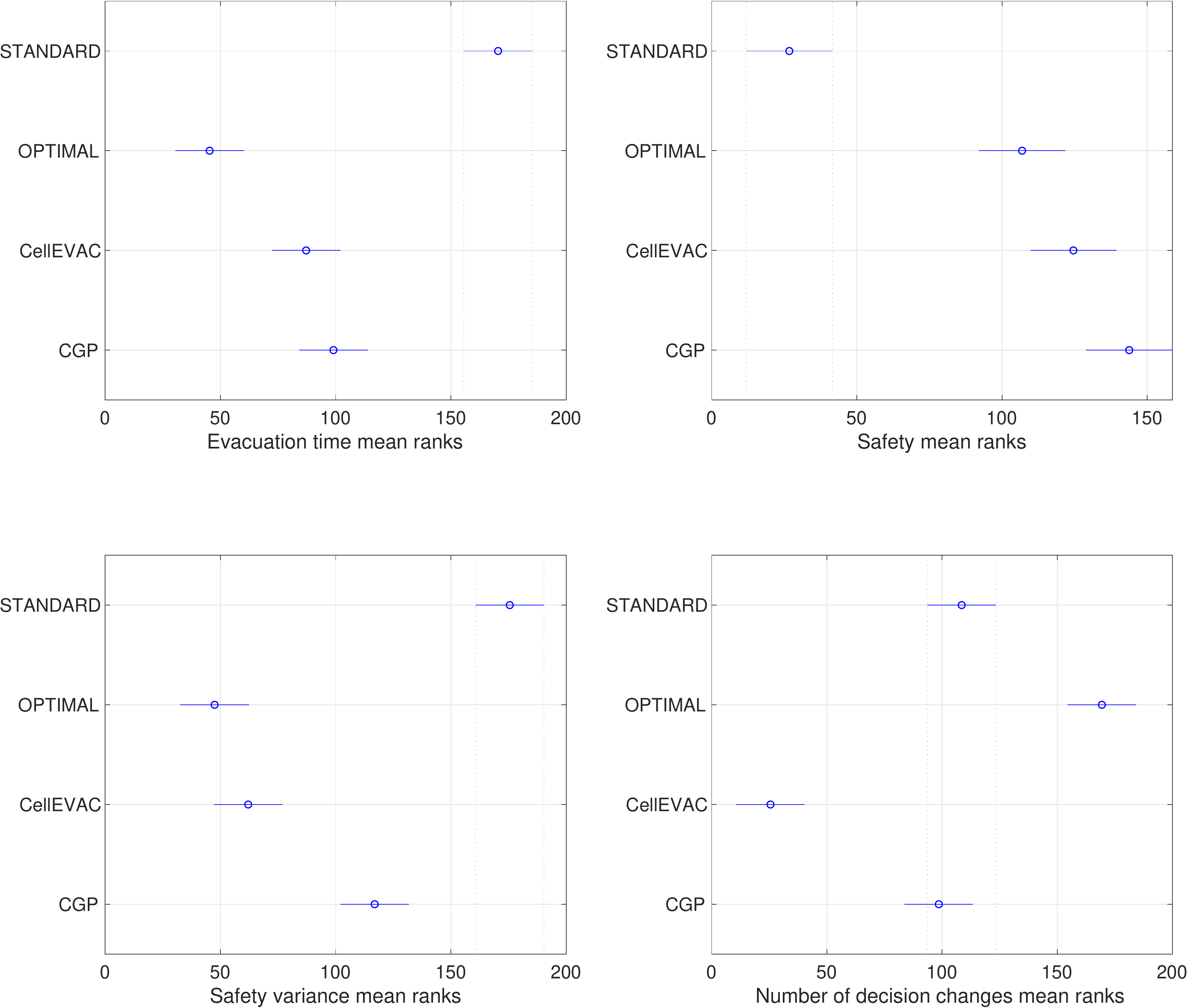}
\par\end{centering}
\caption{\label{fig:Multcompare-General}Multiple comparison tests of the performance
measurements for the STANDARD, OPTIMAL, CellEVAC and CGP configurations.}
\end{figure}

The performance analysis results have been summarized in Figures \ref{fig:Boxplots-General}
and \ref{fig:Multcompare-General}. As expected, the experiments revealed
significantly better overall performance in the optimized models than
the STANDARD behavior. When using the CellEVAC system, the results
were very similar to those obtained when pedestrians followed the
ideal individual strategy defined by the OPTIMAL configuration. The
median evacuation time was around 8 minutes with CellEVAC, only 30
seconds above OPTIMAL. It is interesting to note that the number of
decision changes decreased dramatically with CellEVAC due to the high
degree of coordination imposed by the CellEVAC system through the
exit gate indications. When comparing CellEVAC and CGP, the results
showed a significantly better performance of CellEVAC in terms of
safety variance and the number of decision changes. It confirmed that
CellEVAC balances pedestrian flows significantly better than CGP.

\begin{figure}
\begin{centering}
\includegraphics[width=0.9\columnwidth]{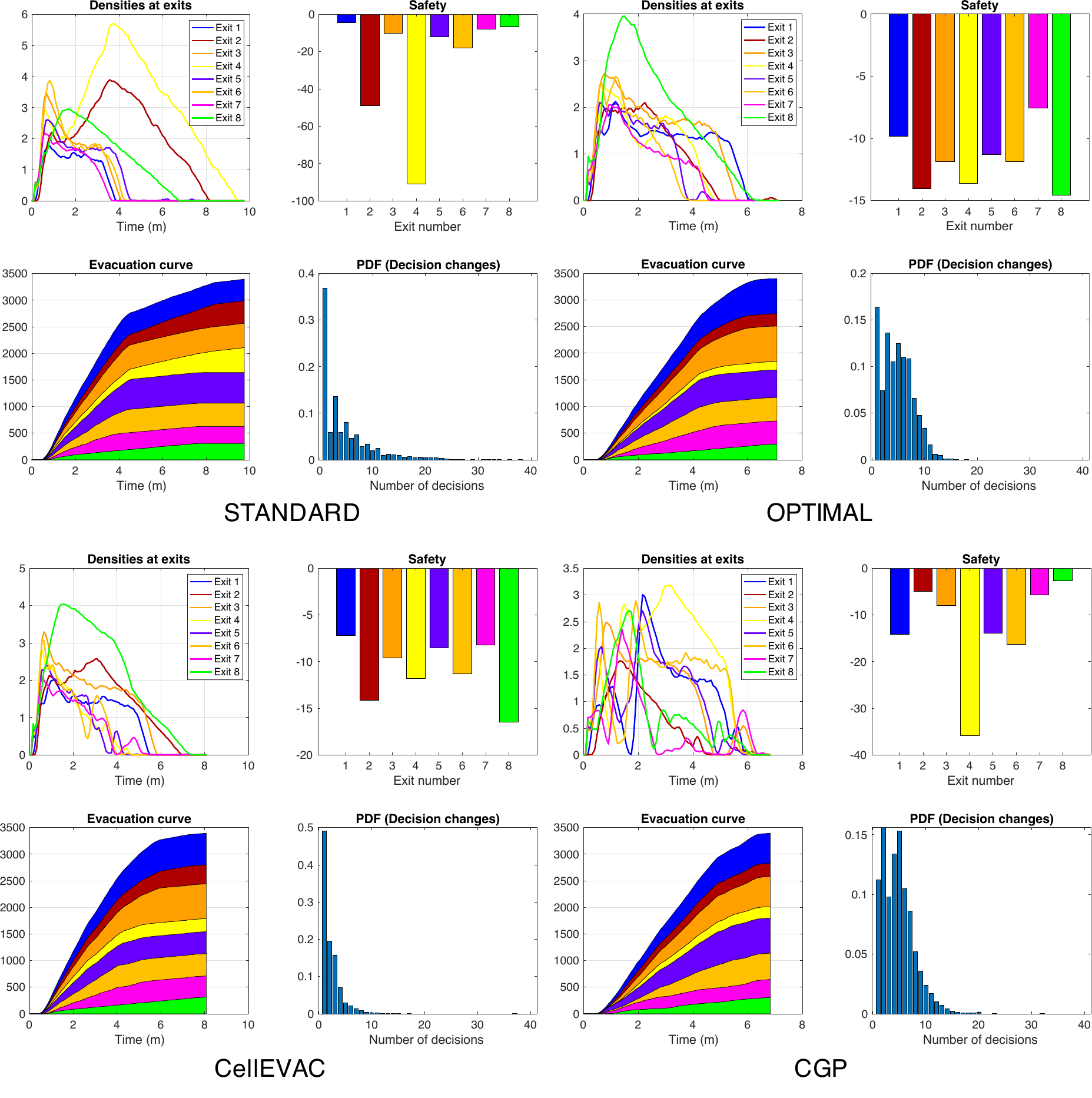}
\par\end{centering}
\caption{\label{fig:Results-of-single-run-behaviour}Results of single run
simulations. Each sub-figure shows the evolution of the pedestrian
densities, safety values and evacuation curves at each exit gate,
and the histogram (probability density function) of the number of
decision changes.}
\end{figure}

The outcomes of single run simulations are shown in Figure \ref{fig:Results-of-single-run-behaviour}.
For comparison, external flows in all simulations were injected at
exits 2 and 4, while the entry at exit gate 8 was blocked. Pedestrians'
speed was artificially reduced by a factor of 100 in a restricted
area at the entrance of the exit gate 8 to simulate that an external
event blocked the exit. 

The density and safety plots for the STANDARD strategies showed highly
unbalanced pedestrian flows at the exit gates. Exit 8, which was the
exit with the highest capacity, was underutilized. Also, we observed
low safety levels at exit gates 2 and 4 due to external flows. In
contrast, CellEVAC exhibited balanced pedestrian flows at the exit
gates and significantly lower decision change values compared to the
other configurations. The CGP system showed highly unbalanced pedestrian
flows at the exit gates and significant oscillations in the density
curves. In the same scenario, CellEVAC presented a much more homogeneous
and stable evolution of pedestrian flows. Notably, the distribution
of decision change values showed an inferior performance in CGP, significantly
skewed to the right.

\begin{figure}
\begin{centering}
\includegraphics[width=0.8\columnwidth]{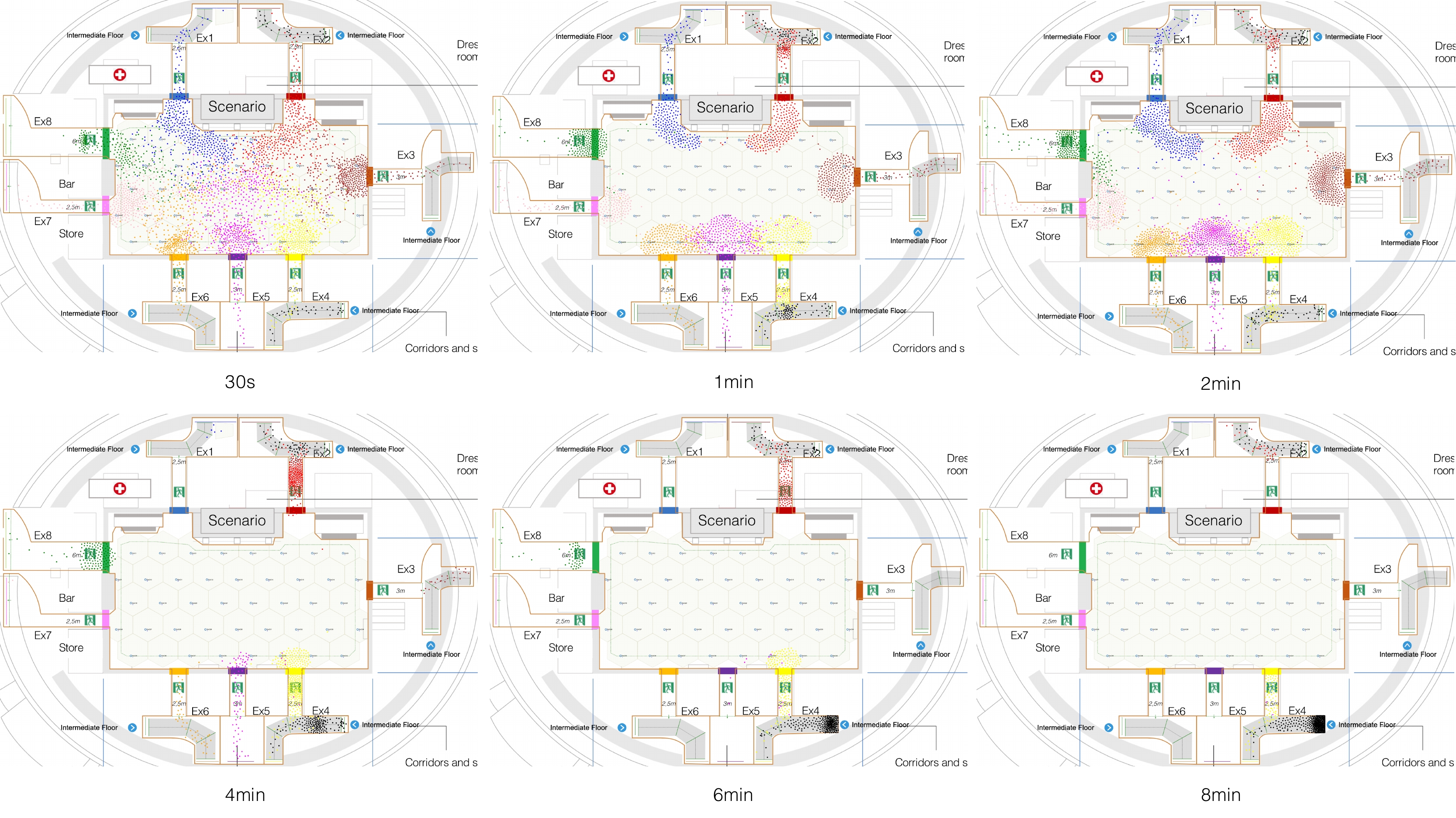}
\par\end{centering}
\caption{\label{fig:Still-images-from-behaviour-MLM-STANDARD}Still images
from the single run simulation experiments of the STANDARD behavior.}
\end{figure}

Still images of the single run simulation experiment for the STANDARD
configuration in Figure \ref{fig:Still-images-from-behaviour-MLM-STANDARD}
showed the accumulation of pedestrians at both exits. With STANDARD,
it can be observed how exits 2 and 4 (red and yellow) are still collapsed
at minute 6, while with OPTIMAL exits 1 and 5 share their capacity
to discharge exits 2 and 4 respectively (Figure \ref{fig:Still-images-from-behaviour-MLM-OTSEF}).
Contrary to expectations, safety at exit 8 was high, but the explanation
is that its capacity doubles the capacity of the remaining exit gates.
It is crucial to note that with the optimal strategy the balance of
the pedestrians flows at the exit gates is significantly improved,
and so the safety and evacuation time. These results exemplify the
importance of the right balance of pedestrian flows in the improvement
of evacuation processes. The simulations revealed that OPTIMAL provided
a much better balance between the exit gates at the cost of significantly
higher decision changes (histograms in Figure \ref{fig:Results-of-single-run-behaviour}). 

\begin{center}
\begin{figure}
\begin{centering}
\includegraphics[width=0.8\columnwidth]{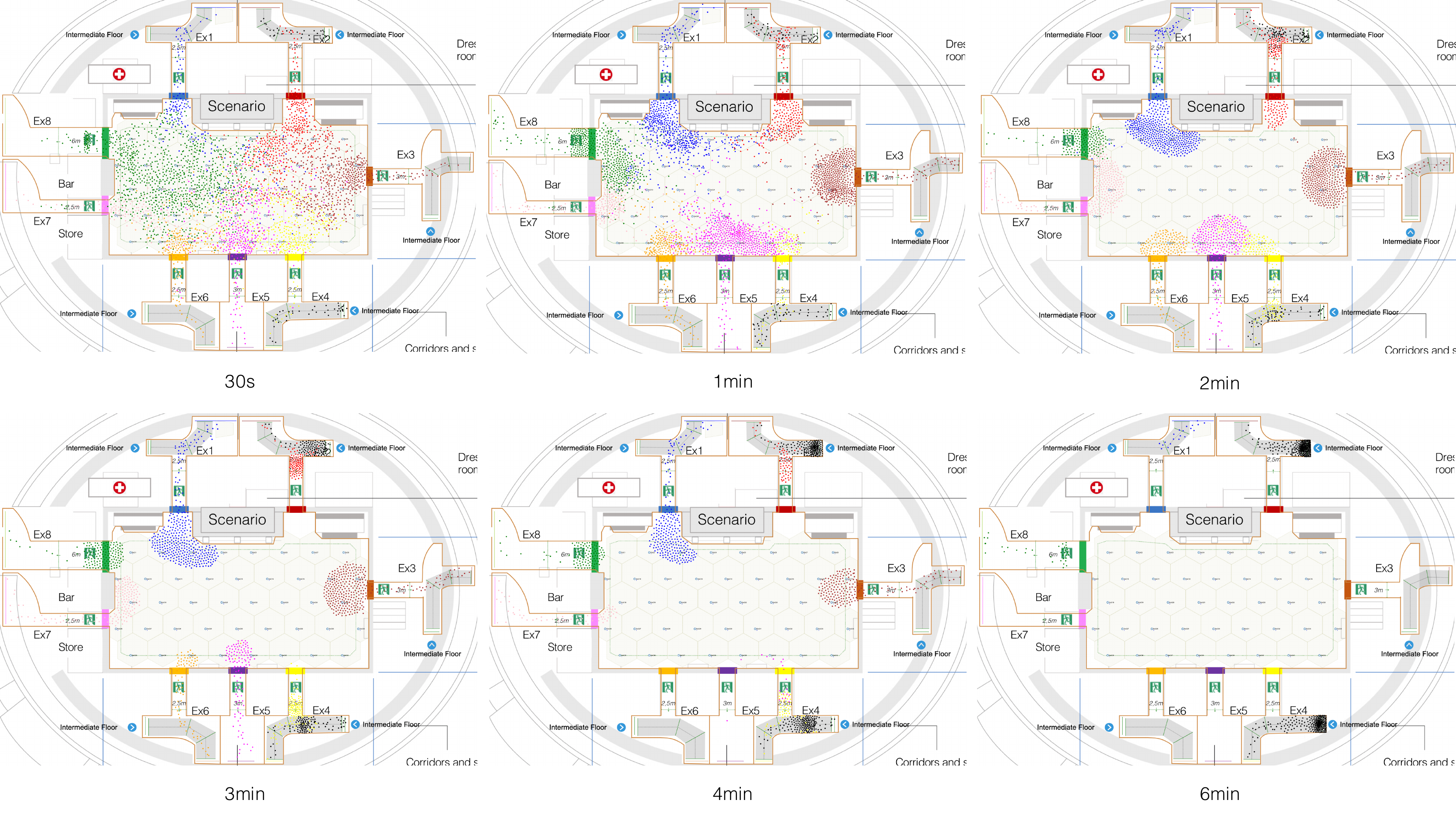}
\par\end{centering}
\caption{\label{fig:Still-images-from-behaviour-MLM-OTSEF}Still images from
the single run simulation experiments of the OPTIMAL individual behavior.}
\end{figure}
\par\end{center}

\begin{figure}
\begin{centering}
\includegraphics[width=0.6\columnwidth]{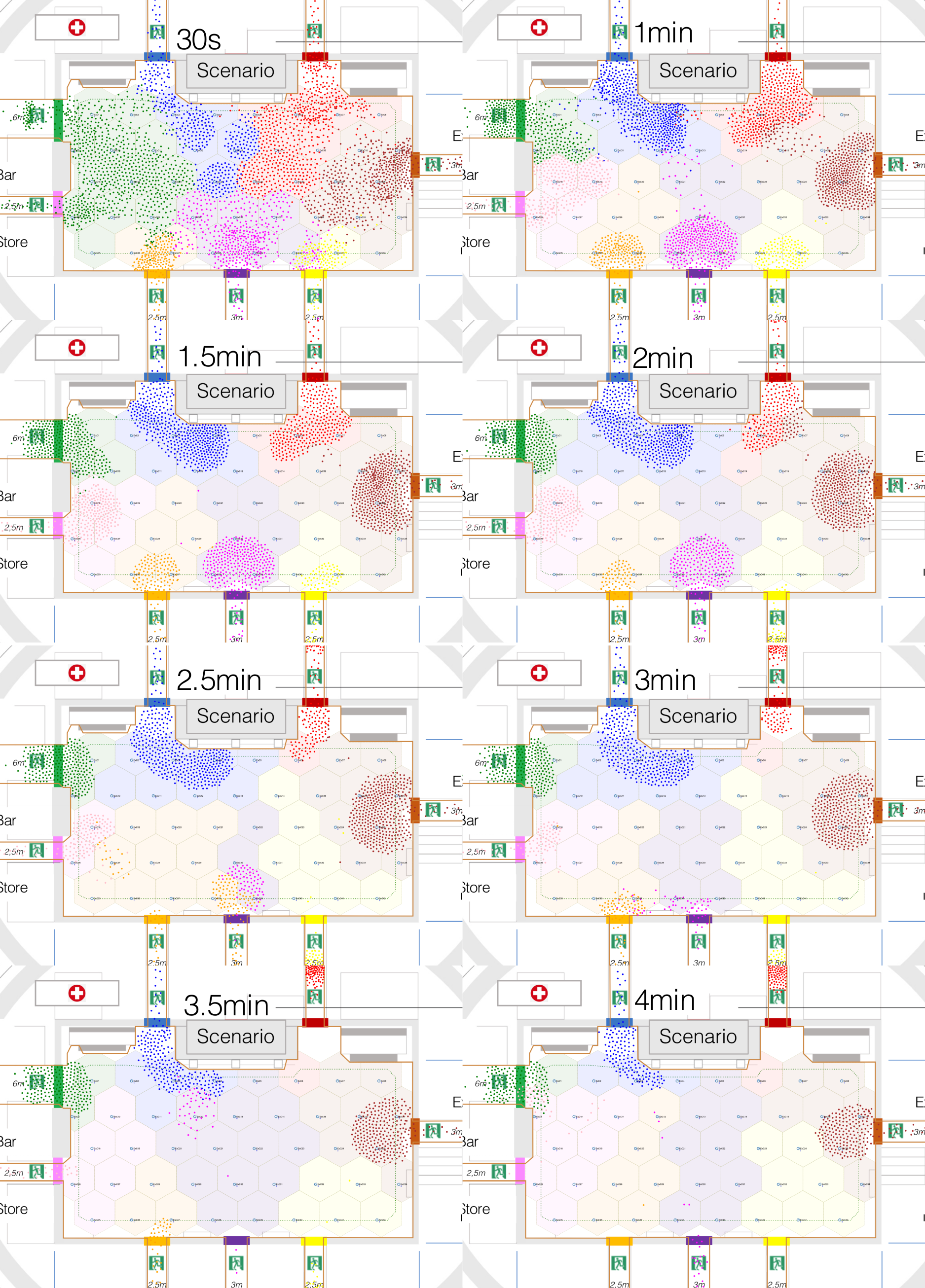}
\par\end{centering}
\caption{\label{fig:OTSEF-EF-CELL-still-images}Still images from the single
run simulation experiments of the CellEVAC system with external flows
at exit gates 2 and 4, and exit gate 8 blocked. Each cell shows the
color corresponding to the recommended exit gate.}
\end{figure}

In Figure \ref{fig:OTSEF-EF-CELL-still-images}, still images of the
evacuation experiment when using CellEVAC showed how exits 1, 5, and
7 (blue, magenta, and pink) shared their capacity to discharge the
congested exits 2, 4 and 8 (red, yellow and green) respectively. It
is important to note that pedestrians' movement is more homogeneous
when using the CellEVAC guidance system, compared to pedestrians making
decisions at the individual level without following indications (see
Figure \ref{fig:Still-images-from-behaviour-MLM-STANDARD}). Overall,
these results have further strengthened our confidence in the CellEVAC
MLM system as an effective adaptive guidance system.

\begin{figure}
\begin{centering}
\includegraphics[width=0.6\columnwidth]{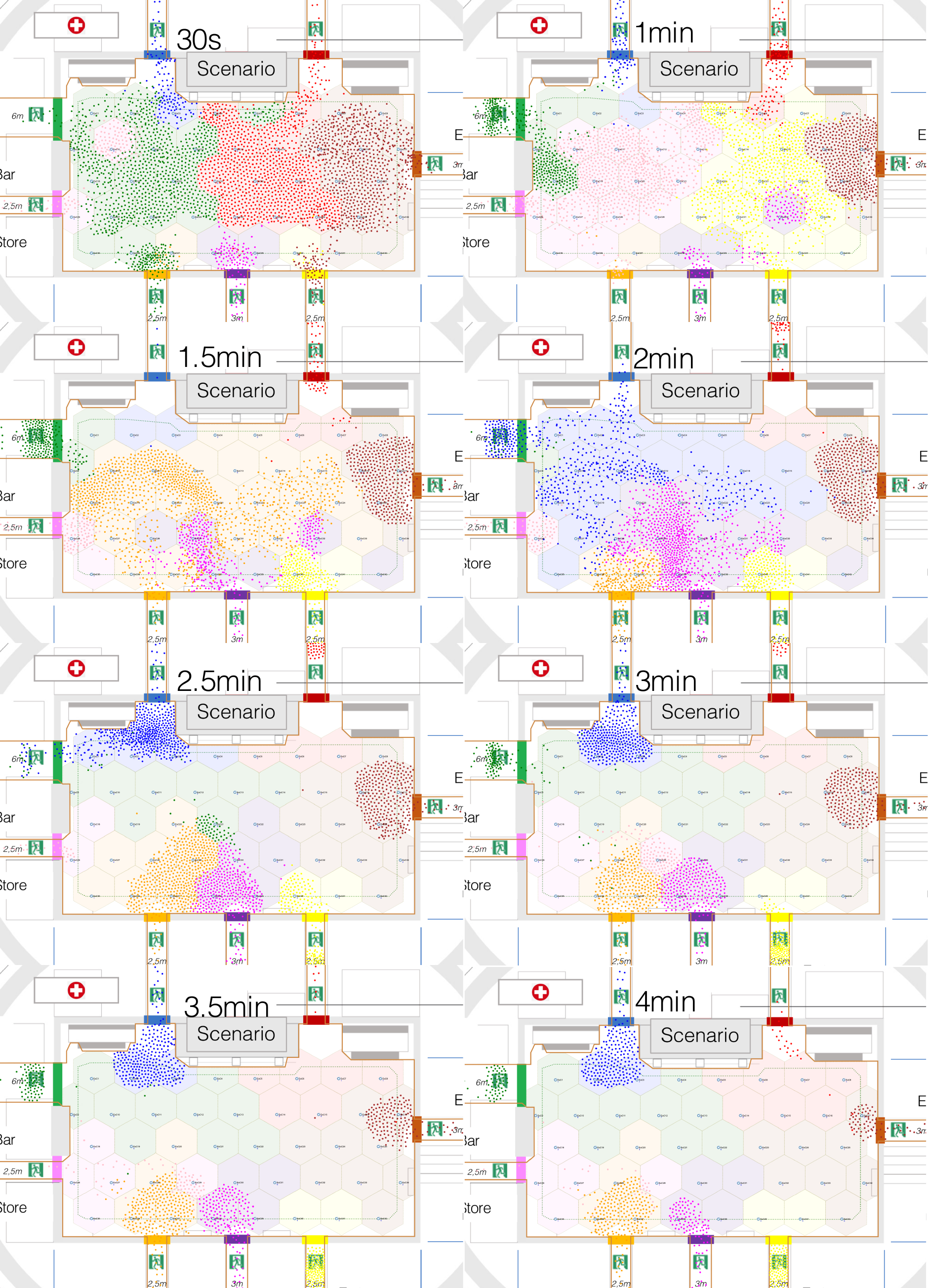}
\par\end{centering}
\caption{\label{fig:OTSEF-EF-CGP-still-images}Still images from the single
run simulation experiments of the CGP system with external flows at
exit gates 2 and 4, and exit gate 8 blocked. Each cell shows the color
corresponding to the recommended exit gate.}
\end{figure}

Finally, still images for the CGP single run experiment in Figure
\ref{fig:OTSEF-EF-CGP-still-images} showed how different exit gates
discharged the congested exit gates 2 and 8 (red and green) at different
moments. However, exit gate 4 (yellow) remained congested during the
evacuation process. In contrast, with CellEVAC, the three congested
exit gates were discharged by adjacent exit gates with available capacity,
exhibiting a much more stable behavior. It is important to note that
pedestrians' movement suffered from large oscillations, showing a
highly unnatural behavior. 

Using an optimized behavior model to implement the decision logic
of a guidance system, as hypothesized, helps to provide adequate control
actions (indications) more similar to human behavior. We believe that
this kind of indication will be easier to follow by pedestrians in
evacuation scenarios characterized by high uncertainty, stress, and
panic.

\subsection{How the compliance rate of CellEVAC influences evacuation efficiency}

\begin{figure}
\begin{centering}
\includegraphics[width=0.6\columnwidth]{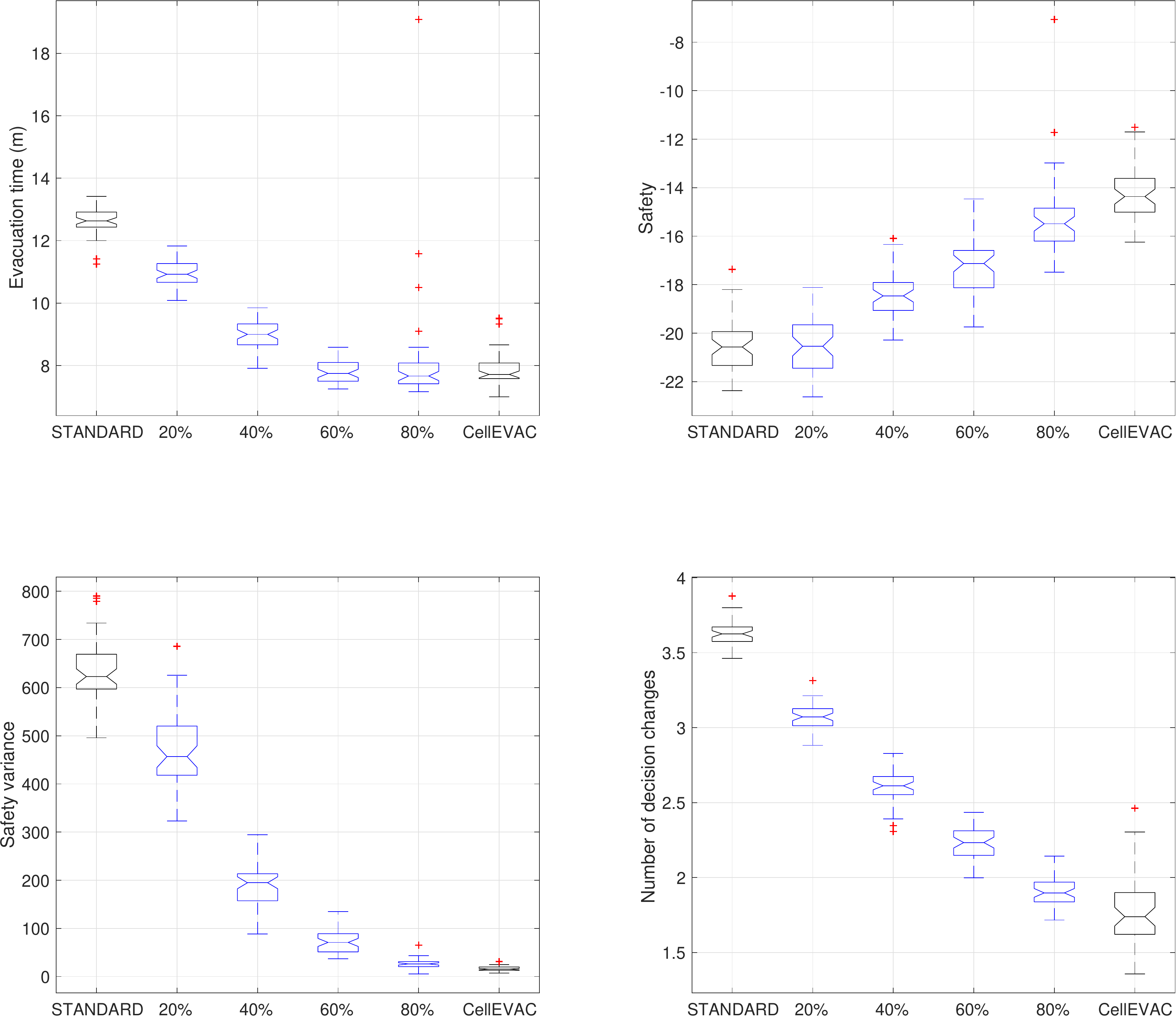}
\par\end{centering}
\caption{\label{fig:Boxplots-MIXED}Box-plots of the sensitivity analyses of
the performance measurements to the compliance rate of CellEVAC. Horizontal
axes categorize the compliance rate from 0\% (STANDARD) to 100\% (CellEVAC).}
\end{figure}

The outcomes of the CellEVAC sensitivity analyses to compliance rate
are presented in Figure \ref{fig:Boxplots-MIXED}. Compliance rate
ranged from 0\% (STANDARD) to 100\% (CellEVAC) in increments of 20\%.
The percentage of users using CellEVAC were committed to follow indications
during all the evacuation process, while the remaining users followed
the STANDARD behavior. 

Based on sensitivity analyses to evacuation time, the system required
60\% of people to use CellEVAC to achieve the best value of evacuation
time. As regards average safety and safety variance, results showed
a linear improvement. Finally, the number of decision changes also
exhibited a linear improvement with an increasing compliance rate.
Overall, the most remarkable result of the sensitivity analyses is
that the system is highly effective, even at low compliance rates.

\subsection{\label{sec:Discussion}Discussion}

Our use of an optimized exit-choice behavior model to implement the
control logic of an adaptive guidance system of exit-choice has proven
to be very efficient in terms of the evacuation time, safety, and
the number of decision changes. This technique improves existing non-adaptive
approaches, which are very efficient in evacuation time but generate
fixed plans incapable of responding to unexpected conditions \citep{Abdelghany20141105,Zhong20141125,wongOptimizedEvacuationRoute2017}.
The experimental evaluation confirms that CellEVAC performs better
than adaptive proposals based on heuristic rules (CGP) \citep{Zhong2016127}.
Moreover, our experimentation with CGP was challenging concerning
the optimization process. 

The operation of CellEVAC confirms the hypothesis of a much more natural
response to exit-choice indications than those obtained with heuristic
rules. In evacuation scenarios, characterized by high levels of stress,
uncertainty, and panic, receiving instructions that are optimal and
closer to human behavior patterns are particularly important for a
much higher utilization rate and efficiency. Significantly, CellEVAC
proves to be effective even in scenarios with a utilization rate of
40\%. Imitation behavior seems to be crucial in its effectiveness
in this regard.

Up to our knowledge, this is the first evacuation guidance system
architecture based on color codes for the adaptive recommendation
of exit gates. We strongly believe that this architecture has a high
potential for its simplicity. From the physical deployment perspective,
its cost and complexity may be small, as explained in the architecture
description. Besides, the human-machine interface provides clear and
simple instructions that are easy to follow. This latter aspect complements
the natural behavior patterns generated by the exit indications given
by CellEVAC.

Another of our main goals was to include the safety measure as a performance
objective. The result is that the optimization processes search for
the right balance of the pedestrian flows using the modeled dynamic
response of the exit gates throughout pedestrian fundamental diagrams.
This approach has allowed improving the evacuation processes significantly.
Note, however, that the measurement of safety in our research depends
on arbitrary density thresholds for comparison purposes. Depending
on the specific scenario, and the forecast of capacity and flows,
it would be necessary to apply some calibration technique to set the
adequate thresholds in the pedestrian fundamental diagrams. 

We are aware that the average and variance of safety at the exit gates,
could be replaced by other statistics depending on the specific application,
and the analysis and calibration based on real experiments. For example,
instead of using the average value, we could be more conservative
using a maxi-min objective function. Finally, we have considered safety
at exit gates, ignoring the analysis of safety in other areas. In
an evacuation scenario with a relatively small size as the one addressed
in our work, it seems reasonable to focus the study at the exit gates,
which are typically the most dangerous zones. However, in larger scenarios,
it would be interesting to extend the safety measurements to the entire
evacuation area.

In line with \citep{haghaniSimulatingDynamicsAdaptive2019}, modeling
the exit-choice change has proven to be a critical parameter in modeling
optimal individual behavior and modeling and optimizing the CellEVAC
decision control module. The integration of exit-choice changing behavior
and the remaining attributes of exit-choice decision making, allowed
us to significantly reduce the number of degrees of freedom of the
model and simplify the optimization process. However, this does not
exclude the study of alternatives based on a two-layered model.

Interestingly, the results highlight the importance of imitation patterns
in individual behavior, which are nevertheless of sign negative in
the CellEVAC system. The grouping effect of CellEVAC indications seems
to inherently incorporate the imitation pattern, which has to be compensated
by the negative value of the parameter $\beta_{P}$.

Although the simulation-optimization modeling framework proposed in
this paper is extensible to scenarios different from Madrid Arena,
some aspects are specific and restrict its direct application. For
instance, a different facility will require a different cell structure
depending on its size and shape, maybe with cells of different sizes.
Moreover, we will need to compute the fundamental diagrams of the
exit gates and calibrate their density thresholds. The direct vision
to all the exit gates and the absence of obstacles simplified the
design of the exit-choice models and the infrastructure of light indicator
panels. In scenarios without direct vision and obstacles, we would
need intermediate light indicator panels, in addition to those located
at the exit gates. This fact represents a significant challenge to
investigate.

Regarding the simulation-optimization processes, we have followed
a cross-validation method, in which two different models, for scenarios
with and without external flows, are validated in scenarios with and
without external flows. Another possibility would have been to generate
a single optimal model by using many different pedestrian flow patterns.
However, our objective was not to generalize but to investigate the
influence of different factors on evacuation processes (e.g., different
fitness functions or different expected pedestrian flows). According
to the experimental results, it seems reasonable that by following
the same procedures, it is possible to configure CellEVAC in very
different scenarios.

\section{Conclusions}

We have proposed an adaptive guidance system named CellEVAC for crowd
evacuations based on exit-choice indications. These indications have
the form of colors displayed in personal or wearable devices that
allow evacuees to find the exit gate with the corresponding colored
light indicator panel. This type of indication simplifies greatly
its interpretation, which is particularly important in stressful situations
found typically in evacuation scenarios.

Our research focused on Madrid Arena, an indoor arena located in the
city of Madrid (Spain). We have defined a system architecture that
divides the facility into homogeneous cells such that evacuees in
the same cell receive the same indications. The architecture assumes
an indoor or outdoor positioning system that provides pedestrians
location-aware capabilities. The core of the system is the controller
module (decision logic module) that monitors the environment, decides
on the allocation of exit gates (colors) to cells, and sends this
allocation to sensor nodes located in the center of the cells. This
information is transmitted periodically throughout a broadcast communication
channel. 

The first aim of our study was to assess the use of a pedestrians'
exit-choice behavior model to implement the decision logic module.
We built this module upon the simplest and most popular practical
discrete choice model, the Multinomial Logit Model (MLM). Our goal
was to find the optimal configuration of the model to optimize evacuation
time and safety. Thus, the procedure for conducting the optimization
processes was to adopt a simulation-optimization approach in which
heuristic search algorithms integrate with a microscopic pedestrian
simulation based on the Social Force Model. 

We built two different simulation-optimization frameworks integrating
pedestrian behavior modeling, social force model for pedestrian motion,
control logic of exit gate indications, and optimization features.
In the first framework, AnyLogic simulation software interconnected
with a Matlab engine. We used this software framework to perform Tabu-search
optimization of the MLM model and perform sensitivity and performance
analyses. The second software framework interconnected AnyLogic with
a Matlab engine and ECJ (Evolutionary Java Computation Framework).
This configuration was used to optimize, evaluate, and compare an
existing control logic based on Cartesian Genetic Programming. With
the interconnection of these programming, simulation, and optimization
software applications, we have found a cutting-edge solution for developing
simulation-optimization in the field of pedestrian evacuation.

The second aim of our research was about defining a safety metric
for evacuation processes to be used in the optimization processes
as an explicit objective. Something neglected in the existing literature.
We have introduced a method in which we first calculate the pedestrian
fundamental diagrams of the exit gates to capture their dynamics.
Based on the fundamental diagrams, the next step is to establish specific
density thresholds for each exit gate that determine what is considered
safe. Following this procedure, we have described overall statistics
that define performance safety values systematically and objectively.

With these two main objectives in mind, we first investigated how
evacuation performance could be influenced by different individual
behaviors, paying attention to exit-choice changing strategies. The
sensitivity analyses and optimization results have underlined the
importance of imitation behaviors and exit-choice changing modeling
in the performance of evacuation processes. In contrast to existing
research, we incorporated the modeling of exit-choice changing behavior
in the exit-choice decision model. We have confirmed the viability
of this integration and the simplification of the optimization processes.

Next, we moved on to the optimization tasks at a system level. Our
goal here was to optimize the MLM behavior model used in the decision
logic module of CellEVAC. We obtained the best results when the optimization
objective was defined explicitly in terms of evacuation time and safety,
and the evacuation scenario included complex pedestrian flows. The
performance analyses of the model obtained, confirmed a balanced pedestrian
flow and a natural movement to the exit gates, in addition to a similar
performance when compared to the optimal behavior at an individual
level. Interestingly, imitation behavior disappeared from the optimal
MLM model in CellEVAC due to the grouping effect inherent to the CellEVAC
indications. 

We compared the MLM behavior model with an existing Cartesian Genetic
Programming approach based on the optimization of heuristic rules.
The results confirmed the advantage of using a behavior model to control
the indications of exit-gates. The use of heuristic rules exhibited
a worst overall performance, with unnatural movements and an excessive
number of decision changes. Moreover, we found it quite problematic
to configure the optimization process.

The evidence from this study suggests the viability of CellEVAC as
an effective adaptive guidance system based on exit-choice indications.
Taken together, the results confirm that optimized pedestrian behavior
models can be effectively used to develop decision logic modules in
adaptive guidance systems for crowd evacuations.

Several extensions are considered for this research. We are in the
process of investigating specific technologies to implement and deploy
the CellEVAC. Also, we will examine the differences between a single-layered
decision model and a two-layered model with separate functions for
exit-choice changing and exit-choice decision. We will need to investigate
how to reduce the complexity of the optimization processes with two-layered
models. Effort is also underway to study the use of CellEVAC in different
evacuation scenarios with obstacles and dynamic conditions in the
facility. Another research extension is related to the calibration
with real data of the safety model. In a deployment of CellEVAC in
real scenarios, we will need to determine the optimal density thresholds
to improve safety. Finally, in mega facilities such as grand stadiums,
developing a safety metric that includes the pedestrian flow dynamics
of all the facility and not the exit gates only, will need to be undertaken.

\section*{Acknowledgements}

This work was supported by the Spanish Ministry of Economy and Competitiveness
under Grant TIN2016-80622-P. The authors are most grateful to all
the reviewers for their comments and recommendations on the text.

\bibliographystyle{model5-names}

\appendix
\setcounter{figure}{0} \renewcommand{\thefigure}{A.\arabic{figure}}

\section{\label{sec:Optimization-processes}Optimization processes}

\begin{figure}
\begin{centering}
\includegraphics[width=0.9\columnwidth]{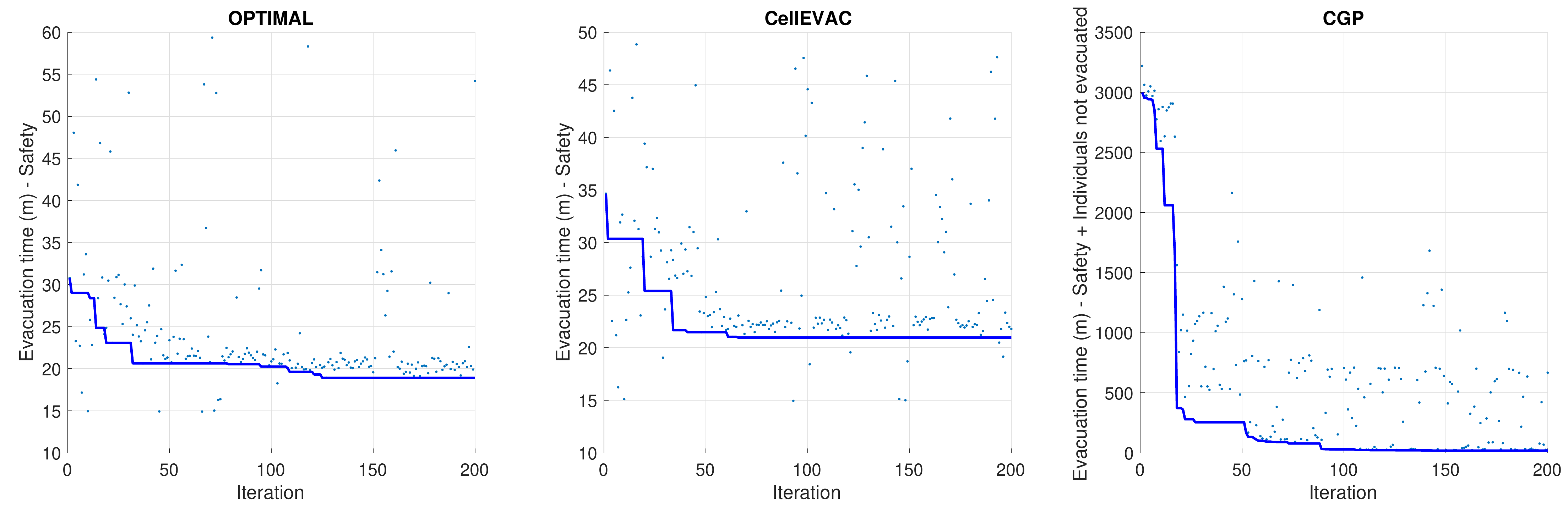}
\par\end{centering}
\caption{\label{fig:Optimization progress}Progress of the simulation-optimization
processes. For OPTIMAL and CellEVAC the blue dots below the current
best-solution line represent the non-viable solutions obtained during
the Tabu-search optimization process.}
\end{figure}

Figure \ref{fig:Optimization progress} illustrates the progress of
the simulation-optimization processes to obtain the optimal configurations
for the OPTIMAL, CellEVAC and CGP models. We imposed an arbitrary
simulation stop-limit of 15 minutes to evacuation time, after which
the simulation iteration stopped. The aim was to avoid the consumption
of simulation time in non-viable solutions during the optimization
process. 

In the optimization of the individual pedestrian behavior (i.e., OPTIMAL
configuration) and the CellEVAC system, a restriction to the viability
of solutions was incorporated in the Tabu-search algorithm, removing
solutions in which there were pedestrians pending evacuation. We discovered
that the search process could find solutions in which pedestrians
did not move, artificially increasing the objective value due to an
average safety close to 0.

For the CellEVAC guidance system, we assumed that the entire population
of evacuees followed the indications of the CellEVAC system, and therefore
in the pedestrians' MLM behavior model, all the parameters were equal
to 0, except for $\beta_{SYS}=1$. 

The simulation-optimization process of the CGP model was configured
following the recommendations of \citep{Miller2011} as follows: population
of $1+\lambda=5$ (i.e., population of 5 candidate solutions per iteration
of the genetic algorithm), a mutation rate $\mu_{r}=0.75\%$, $4000$
nodes, $5$ input nodes, and $1$ output node. We found that the algorithm
did not converge to viable solutions even with short genotypes. The
solution was to include in the fitness function, the number of people
waiting to be evacuated at the end of the simulation deadline (15
minutes): $min(evacTime-Sf+numberOfWaitPeds)$. Thus, at the first
iterations of the optimization process the search was directed towards
minimizing pedestrians' number. Once the search process found solutions
that were able to evacuate all the population, the next iterations
were automatically focused on improving evacuation time and safety.
In addition to this modification of the fitness function, we had to
progressively increase the length of the chromosome from 50 to 4000
nodes until we finally found viable solutions. 

The best solution curve shows how fitness values progressively decreased
from 3000 to 20. As in the optimization of CellEVAC, we assumed that
the entire population of evacuees followed the indications of the
CGP module.
\end{document}